\documentclass[prx,reprint,noshowpacs,superscriptaddress,floatfix,letterpaper,longbibliography]{revtex4-2}
\usepackage{amsmath,amssymb,amsbsy,amsfonts,amsthm,mathrsfs,amssymb}
\usepackage{bm,bbm,graphicx}
\usepackage{float}
\usepackage{epsfig}
\usepackage{color}
\usepackage[dvipsnames]{xcolor}
\usepackage[colorlinks=true,citecolor=Cerulean,linkcolor=RubineRed,urlcolor=Cerulean]{hyperref}
\usepackage{silence}
\usepackage{soul}
\usepackage[normalem]{ulem}
\usepackage{xfrac}
\usepackage[version=4]{mhchem}
\usepackage{stackrel}
\usepackage{sidecap}
\sidecaptionvpos{figure}{c}
\usepackage{newfloat}
\DeclareFloatingEnvironment[name={SM Fig.}]{suppfigure}

\DeclareFontFamily{OT1}{pzc}{}
\DeclareFontShape{OT1}{pzc}{m}{it}{<-> s * [0.900] pzcmi7t}{}
\DeclareMathAlphabet{\mathpzc}{OT1}{pzc}{m}{it}

\begin{document}

\title{Universal dynamic scaling in chemical reactions\\at and away from equilibrium}

\author{Shrabani~Mondal}
\thanks{Equal contributions.}
\affiliation{Department of Chemistry,\
 University of Massachusetts Boston,\
 Boston, MA 02125
}
\author{Jonah~S.~Greenberg}
\thanks{Equal contributions.}
\affiliation{Department of Chemistry,\
 University of Massachusetts Boston,\
 Boston, MA 02125
}
\affiliation{Department of Chemistry,\
  Northwestern University,\
  Evanston, IL 60208
}
\author{Jason~R.~Green}
\email[]{jason.green@umb.edu}
\affiliation{Department of Chemistry,\
  University of Massachusetts Boston,\
  Boston, MA 02125
}
\affiliation{Department of Physics,\
  University of Massachusetts Boston,\
  Boston, MA 02125
}
\affiliation{Center for Quantum and Nonequilibrium Systems,\
  University of Massachusetts Boston,\
  Boston, MA 02125
}

\begin{abstract}

Physical kinetic roughening processes are well known to exhibit universal
scaling of observables that fluctuate in space and time.
Are there analogous dynamic scaling laws that are unique to the chemical reaction mechanisms available synthetically and occurring naturally?
Here, we formulate two complementary approaches to the dynamic scaling of stochastic fluctuations in thermodynamic observables at and away from equilibrium.
Both analytical expressions and numerical simulations confirm our dynamic scaling ans{\"a}tze with their associated exponents, functions, and laws.
A survey of common chemical mechanisms reveals classes that organize according to the molecularity of the reactions involved, the nature of the reaction vessel and external reservoirs, (non)equilibrium conditions, and the extent of autocatalysis in the reaction network.
Coupled reactions capable of chemical feedback can transition, sometimes sharply, between these classes with the variation of experimental parameters such as temperature.
While path observables like the dynamical activity have scaling exponents that are time-independent, fluctuations in the entropy production and flow can have time-dependent scaling exponents and self-averaging properties as a result of temporal correlations that emerge during thermodynamically irreversible processes.
Altogether, these results establish dynamic universality in the nonequilibrium fluctuations of thermodynamic observables for well-mixed chemical reactions.

\bigskip

\hfill%
\begin{minipage}{12cm}
{\footnotesize Subject Areas: Nonequilibrium Statistical Mechanics, Dynamic Scaling, Chemical Kinetics}
\end{minipage}

\end{abstract}

\maketitle

\section{Introduction}

Chemical reaction mechanisms have the functionality and the diversity to create
materials, synthesize medications, and sustain life.
These kinetic mechanisms can be difficult to classify, however, in part because of this diversity and, in part, because of their nonequilibrium nature.
By contrast, it is well known in statistical physics that seemingly unrelated physical phenomena, from sandpiles to earthquakes, can share universal laws when we change the time and length scale of our observation~\cite{Barabasi1995,Vicsek92}.
Moreover, simulations of and experiments~\cite{buldyrev92,balankin2006,miranda_2010,Takeuchi2011} on growing interfaces have shown that concepts of scaling and universality can apply beyond equilibrium critical phenomena to systems driven out of equilibrium~\cite{Odor2004}.
Despite this progress for physical phenomena, it is unclear whether there are complementary \textit{dynamic} scaling laws for chemical reactions at and away from equilibrium.

Universal scaling behavior has been found in biochemical networks~\cite{Kimeaau0149}, the stochastic exponential growth and division
of bacterial cells~\cite{IyerCSD2014,IyerWHLBLCCDS2014}, the growth of
human cancers~\cite{perez2020}, and dissipative self-assembly~\cite{makey2020}.
Formal analogies have expanded the scope of kinetic roughening theory~\cite{Halpinhealy1995215,Meakin1993189} even further by treating the fluctuations of mathematical functions as surrogates for the physical interface~\cite{Barabasi1995}.
Examples include biological systems such as DNA~\cite{Peng1992}, complex networks~\cite{barabasi2013}, crude oil prices~\cite{balankin2007dynamic}, heartbeat signals~\cite{Ashkenazy01}, strongly interacting gases~\cite{Fujimoto2020}, and material fracture~\cite{Morel2020}.
Applying this idea to the Lyapunov exponents of dynamical systems, for example, has revealed that the leading Lyapunov vector of extended dissipative dynamical systems~\cite{Pikovsky94,Pikovsky1998,PazoL10,PazoLP16b} falls within the Kardar-Parisi-Zhang universality class~\cite{Kardar1986}.
Hamiltonian dynamical systems show anomalous non-KPZ behavior; long-range correlations can cause the fluctuations in finite-time Lyapunov exponents in these systems to self-average weakly~\cite{DasG17,DasG2019} and even diverge~\cite{PazoLP16}.
These results highlight not only the breadth of kinetic roughening theory but also how dynamical mechanisms can influence universal behavior.
Given the diversity of chemical transformations, they also motivate a deeper look at the dynamic scaling and potential universality classes of chemical reactions.

Chemical reactions are dynamic phenomena involving transformations of energy, which makes the fluctuating observables of stochastic thermodynamics good candidates for an analogy with surface roughening.
In the framework of stochastic thermodynamics~\cite{Jarzynski11,seifert2012stochastic,Vandenbroeck2015}, quantities, such as heat, work, and entropy, can be treated at the level of individual, fluctuating trajectories.
Nonequilibrium fluctuations are known to obey strong relations, including fluctuation theorems and thermodynamic uncertainty relations~\cite{Marsland_2017,Seifert2018,horowitz2019thermodynamic}, some of which can be cast as thermodynamic speed limits~\cite{ShiraishiFS2018,Ito2018,nicholson2018nonequilibrium,FalascoE2020,Nicholson2020}.
Here, we show the fluctuations of thermodynamic observables also satisfy dynamical scaling hypotheses in well-mixed chemical reaction systems.
We do this by analyzing the nonequilibrium fluctuations in stochastic thermodynamic
observables of a broad set of elementary and coupled chemical reactions evolving at and away from equilibrium.
Numerical and analytical agreement with our scaling hypotheses suggest the universality of fluctuations in dynamical observables.
The corresponding scaling exponents do not depend on the microscopic details of the system or the nature of the chemical species.
In some cases, the exponents are independent of the reactions conditions such as temperature or whether the reaction vessel is open or closed to the flux of matter.
Our approach enables us to divide chemical kinetic processes into classes according to the values of these characteristic scaling exponents.

\section{Models and methods}

\subsection{Model chemical-reaction systems}

To model chemical systems of coupled reactions, we adopt a standard framework
at the mesoscopic level: we consider well-mixed chemical populations of finite
number in a reaction vessel with volume $V$.
Each chemical system we consider is composed of $n$ chemical species $\mathbf{X}:=\{X_1, X_2, \ldots,X_n\}$ with
$X_k$ representing the number of molecules of the $k$th species at a time $t$.
These species can undergo a total of
$m$ reactions, each reaction having a stoichiometric or state-change vector
$\boldsymbol{\nu}_j\in \mathbb{R}^n$ whose $i$th element is the change in the
number of $X_i$ molecules caused by the $j$th reaction.
The evolution of the entire mixture is governed by the chemical master equation~\cite{McQuarrie67}, which for the time evolution of the probability distribution, $P(\mathbf{X},t)$, is:
\begin{equation}
  \frac{dP(\mathbf{X},t)}{dt} = \sum_{j=1}^m\left[a_j(\mathbf{X}-\boldsymbol{\nu}_j)P(\mathbf{X}-\boldsymbol{\nu}_j,t)-a_j(\mathbf{X})P(\mathbf{X},t)\right].
\end{equation}
This equation of motion
can be solved numerically with the finite-state projection
method~\cite{Munsky06,PelesMK06}, thresholding~\cite{nicolaou2020}, and the stochastic simulation algorithm~\cite{press2007numerical}.

To extract scaling laws and scaling exponents, we use stochastic simulations of the chemical kinetics, kinetic Monte Carlo using the Doob-Gillespie algorithm~\cite{Gillespie76,Gillespie77,Gillespie07}.
This algorithm generates an ensemble of realizations that represent the solution to the master equation in the infinite sample limit~\cite{McQuarrie63,*McQuarrie64,Gillespie92}.
Each realization represents the composition (number of molecules of each chemical species) of a mixture of $N$ molecules in a volume $V$ evolving over time.
That is, the mixture advances through a time-ordered sequence of chemical compositions $\hat{\mathbf{X}}(t)$ by way of chemical reaction events with exponentially-distributed waiting times~\cite{Gillespie1991markov}.
In the results that follow, we simulate a wide range of reactions for varying rate parameters, temperature, and initial number of reactants for each reactive system of interest keeping volume, $V$ fixed.

The simulations require a chemical mechanism (the
elementary reaction and their associated rate constants) and experimental
conditions, such as volume, temperature, and initial number of molecules.
Along a stochastic trajectory of the mixture, each reaction occurs in an infinitesimal
time interval $[t,t+dt)$ with probability $a_j[\mathbf{X}(t)]dt$.
The propensity function, $a_j[\mathbf{X}(t)]$, depends on the molecularity $b$ of
the reaction.
For example, unimolecular reactions, A $\to \emptyset$, have a propensity $c_jX_{\text{A}}(t)$ with stochastic rate constant $c_j\propto 1$; bimolecular reactions, A + B $\to \emptyset$, have a propensity $c_jX_{\text{A}}(t)X_{\text{B}}(t)$ with $c_j\propto 1/V$.
The rate constants $k_j$ of each reaction are related to the stochastic rate constant $c_j$ through combinations of the volume and Avogadro's number $N_A$ that depend on the reaction molecularity.
For example, for a second-order reaction, the relation is $c_j=k_j/N_A V$~\cite{Gillespie1991markov}.

\subsection{Stochastic thermodynamics and kinetics}

Fluctuations in several thermodynamic observables fit within the scaling theory
we present here.
For example, the number of configuration changes in the reaction mixture over a given time interval is a common measure of the lability of the dynamics through configuration space (of chemical compositions)~\cite{HedgesJGC09}.
This ``dynamical activity'' for each stochastic trajectory $\hat{\mathbf{X}}_x(t)$ is the number of reactions occurring in a mixture of $N$ molecules over an observation time $t$: $K(x,N,t) := K[\hat{\mathbf{X}}_x(t)]$.
For other model systems, this counting observable has revealed dynamical phase transitions~\cite{Lecomte2007}.
And, even for the well-mixed reaction vessels, we consider here, its distribution over trajectory ensembles is not necessarily Poissonian.

\begin{SCfigure*}[][t]
\centering

\caption{\label{fig:fig1}\footnotesize{\textit{Dynamic scaling for the stochastic chemical kinetics of unimolecular decay, A$\rightarrow$B.} (a) The cumulative reaction count
$K(x,N,t)$ across a representative ensemble of simulated trajectories at seven different times, $t=0.1, 1, 2.5, 5, 10, 15, 25$ in $1/c$ units.
Mean, $h_K(N,t)$, and fluctuations in the
reaction count, $w_K^2(N,t)$, grow as $t^\delta$ with $\delta = 1$ up to the crossover at time $t_\times = c^{-1}\ln 2$ after which the mean saturates and the fluctuations decay to zero.
For a given $c$, data for the (b) mean activity as a function of time collapse onto a single curve when (c) scaled by system size $h_K(N,t)/N^\gamma$ with $\gamma=1$.
For a given $c$, data for the (d) variance $w_K^2(N,t)$ as a function of time also collapse onto a single curve when (e) scaled $w_K^2(N,t)/N^\gamma$ by the system size with $\gamma=1$. Time is scaled $t \to t N^\zeta$ and for this unimolecular reaction the dynamic exponent is $\zeta=0$.
Points are numerical data and dashed lines are the analytical expression.
Colors indicate $\kappa = c = 0.1$ (green), 1.0 (black), and 10.0 (blue) 
with darker colors indicating larger $N$: $N=10^2-10^6$ molecules.
Insets in (c) and (e) show that scaling time by $c$ collapses data for all rate constants.}}

\includegraphics[width=0.6\textwidth,angle=0,clip]{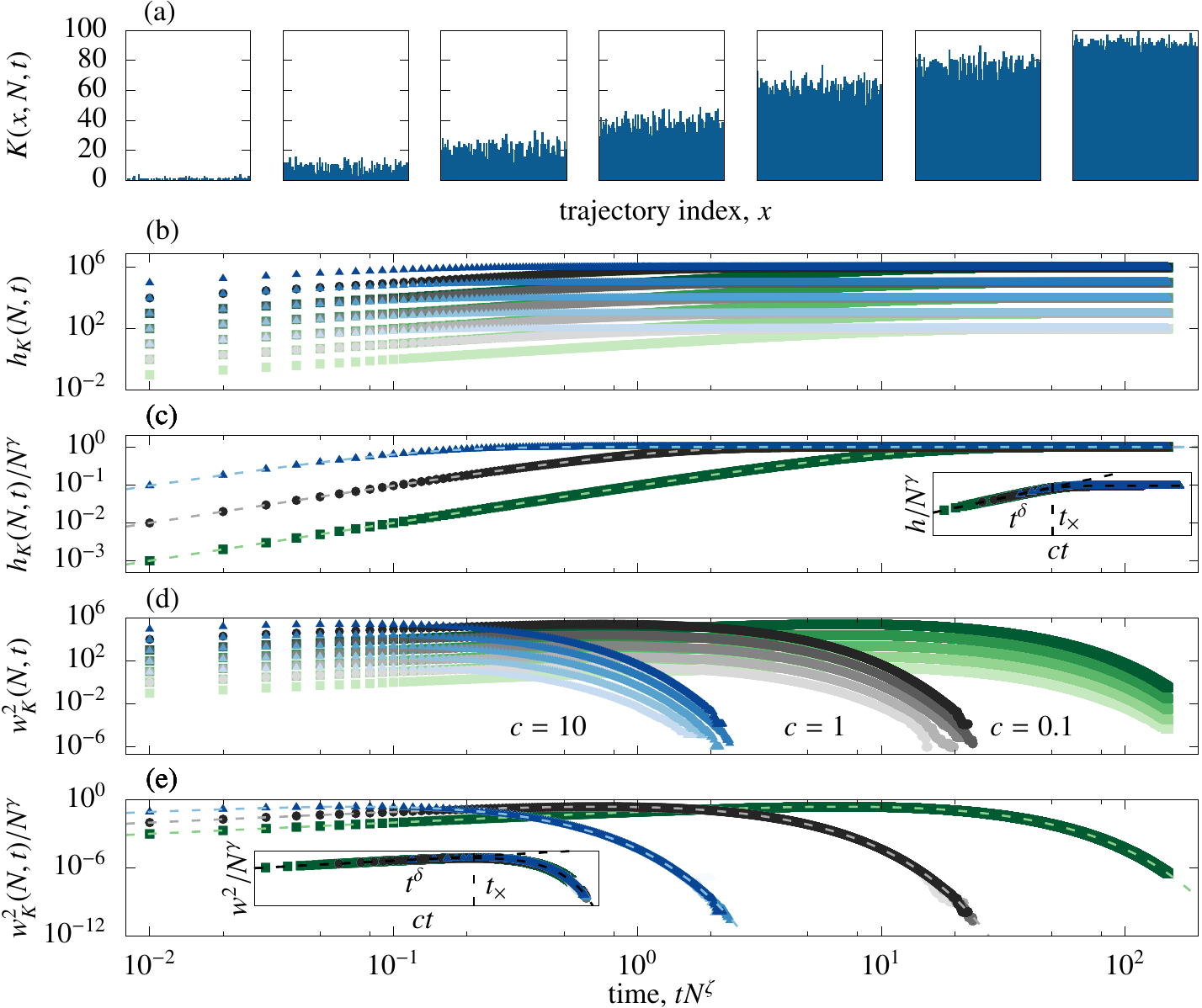}

\end{SCfigure*}

Others have analyzed the stochastic thermodynamics of chemical reaction networks~\cite{mou_stochastic_1986,schmiedl_stochastic_2007,polettini_irreversible_2014,RaoEsposito2016,rao_conservation_2018}.
Complementing this work, we also look at the entropy flow, which for systems that are local detailed balanced, is directly related to the heat dissipated to or absorbed from surroundings.
The action functional~\cite{Lebowitz1999} along the path,
\begin{equation}
 Q_s(x,N,t) = \sum^{K-1}_{i=0} \ln\frac{a_{j}[\mathbf{X}(t_i)\to\mathbf{X}(t_i)+\mathbf{\nu}_j]}{a_{j}[\mathbf{X}(t_i)+\mathbf{\nu}_j\to\mathbf{X}(t_i)]},
 \label{eq:qs}
\end{equation}
is often interpreted as the integrated ``entropy flow'' $-\Delta s_e = Q_s(x,N,t)$ resulting from the exchange of entropy with the surroundings~\cite{seifert2005entropy,schmiedl_stochastic_2007}.
This interpretation follows from the decomposition of the entropy change for the system along the path,
\begin{equation}
  \Delta s = \Delta s_i + \Delta s_e = -\ln \frac{P[\mathbf{X}(t_f)]}{P[\mathbf{X}(t_0)]},
  \label{eq:enpro}
\end{equation}
into contributions from the entropy flow and the ``entropy production'' internal to the system $\Delta{s}_i$~\cite{seifert2005entropy,schmiedl_stochastic_2007}.
We will focus more on the entropy flow here. 
This partitioning also applies to ensemble averages in stochastic thermodynamics where for detailed balanced systems the entropy production, $\Delta S_i$ is non-negative and vanishes at thermodynamic equilibrium, a statement of the second law of thermodynamics, $\Delta S_i\geq 0$~\cite{Vandenbroeck2015}.

The entropy production and flow decompose into observables used in information theory and dynamical systems using the branching observables for forward paths
\begin{equation}
  Q_+(x,N,t)=\sum_{i=0}^{K-1}\ln  \frac{a_j\left[\mathbf{X}(t_i)\to \mathbf{X}(t_i) + \mathbf{\nu}_j\right]}{a\left[\mathbf{X}(t_i)\right]}
  \label{eq:qp}
\end{equation}
and their conjugate reverse
\begin{equation}
  Q_-(x,N,t) = \sum_{i=0}^{K-1} \ln\frac{a_j[\mathbf{X}(t_i)+\mathbf{\nu}_j
\rightarrow
\mathbf{X}(t_{i})]}{a[\mathbf{X}(t_i+\mathbf{\nu}_j)]}+\ln\frac{a[\mathbf{X}(t_{K})]}{a[\mathbf{X}(t_{0})]}.
\label{eq:qm}
\end{equation}
At steady-state, the trajectory-average is related to the dynamical entropy per unit time $h_{\scriptsize{KS}}(N) = -\lim_{t\to\infty}\langle Q_+\rangle/t$ in information theory and the
Kolmogorov-Sinai entropy rate in dynamical systems theory~\cite{gaspard2004time,Lecomte2007}.
These observables measure the degree of branching along a particular forward (reverse) stochastic path.
Together, $-\Delta s_e = Q_s = Q_+-Q_-$~\cite{Lecomte2007} they are the integrated entropy flow, the entropy exchanged between the reaction vessel and the surroundings~\cite{Seifert05}.

To illustrate our approach, Fig.~\ref{fig:fig1}(a) shows the dynamical activity (cumulative number of reactions) for an ensemble of stochastic simulations of unimolecular decay A$\stackrel{c}{\to}$B in a closed container.
Initially, the mixture consists of only A.
The mixture evolves to an absorbing state in which only B is in the reaction vessel on a timescale set by the stochastic rate constant $c^{-1}$.
Here, the value of $K(x,t)$ across the ensemble of trajectories is an abstraction of a rough physical surface.
Clearly, though, the path observables we consider are not extensive in
time for all reactions or reaction conditions and the ``surface'' may not roughen indefinitely.

In what follows, we will draw on the analogy with the statistical evolution of these rough surfaces and analyze the finite-size scaling of the dynamic behavior of fluctuations in the activity, branching observables, and entropy flow/production.
Specifically, we analyze the dependence of these cumulative properties of the $x$th trajectory, $A(x,N,t):=A[\hat{\mathbf{X}}_x(t)]$ on the system size (as measured by the total number of molecules).
Within this conceptual framework, we demonstrate two complementary types of scaling for the mean and variance of these observables for both simple and complex chemical reactions.

\section{Dynamic scaling with intrinsic reaction timescale (Type I)}

\subsection{Dynamic scaling ansatz}

To characterize the statistical evolution of an ensemble of $M$ statistically-independent trajectories, we analyze the average: $h_A(N,t) = \langle A(N,t)\rangle = M^{-1} \sum_x^M A(x,N, t)$.
And to quantify fluctuations about this mean across the trajectory ensemble, we use the variance,
\begin{equation}
  w_A^2(N,t) = M^{-1}\sum_x^M\left[A(x,N,t)-h_A(N, t)\right]^2,
  \label{eq:width}
\end{equation}
which depends on time and the number of $N$ molecules in the volume $V$ at $t=0$.
These moments are the basis for our development of scaling relationships and chemical universality classes.
For a fixed reaction volume that is large compared to molecular length scales,
our dynamic scaling ansatz for the mean $h_A$ and variance $w_A$ of trajectory-level
thermodynamic observables is:
\begin{equation}
\begin{array}{r@{}l}
  h_A(N,t) &{}\sim N^{\gamma} \mathpzc{f}_A(\kappa t N^{\zeta})\\
  w_A^2(N,t) &{}\sim N^\gamma \mathpzc{g}_A(\kappa t N^\zeta).
\end{array}
\label{eqn:ansatz}
\end{equation}
The scaling functions of the mean $\mathpzc{f}_A$ and the variance $\mathpzc{g}_A$ are different in most of the reactions we consider.
Their three characteristic exponents are:

\medskip

\noindent (i) The \textit{wandering exponent} $\gamma$ is a measure of the self-averaging property of the observable~\cite{Milchev1986}.
A $\gamma > 0$ implies that the relative fluctuations of $A$ decay as $w_A^2/h_A^2\sim N^{-\gamma}$ with increasing numbers of molecules for all times.
Most, but not all, of the observables we consider here are system-size extensive
with $\gamma=1$ for their mean and their variance.
Observables that are system-size intensive and self-average, however, will have a wandering exponent of $0$ for their mean and $-1$ for their variance.

\medskip

\noindent (ii) The \textit{dynamic exponent} $\zeta$ is specific to the chosen reaction.
For a given reaction, we define a characteristic timescale as the time between reaction events: $t_c := N/\sum_j a_j(0)$ with the total propensity $N^{-1}\sum_j a_j(0)$ per molecule.
This exponent determines the system size dependence of the characteristic time $t_c \sim 1/\kappa N^\zeta$.
For mechanisms with any number of reactions, provided they have the same molecularity $b$, the exponent is $\zeta = b-1$.
The value of $\zeta$ is the same for both mean and variance.
When $t\approx t_c$, $\mathpzc{f}_A\sim\mathcal{O}(1)$ and $\mathpzc{g}_A\sim\mathcal{O}(1)$ so that $h_A(N,t_c)\sim N^\gamma$ and $w_A^2(N,t_c)\sim N^\gamma$.
The dynamic exponent $\zeta$ accounts for the system size dependence of the time between reaction events; increasing $N$ decreases the time between reactions.

The parameter $\kappa$ is a constant specific to the particular reaction that makes $t/t_c = \kappa t N^\zeta$ dimensionless, App.~\ref{app1}.
For example, it is the stochastic rate constant $\kappa=c$ in the case of unimolecular decay, A$\stackrel{c}{\to}$B.

\medskip

\noindent (iii) The \textit{growth exponent} $\delta$ determines the power law growth of the scaling functions.
We find $\mathpzc{f}_A$ and $\mathpzc{g}_A$ for the mean and variance go as $(\kappa t)^\delta$ for the activity, $-Q_+$, and $-Q_-$.
These scaling functions hold at all times for reactive systems at equilibrium and nonequilibrium stationary states and at early times for systems transiently relaxing to stationary states.
The well-mixed chemical systems here all have $\delta=1$ for the mean and variance, reflecting initial Poisson growth.

\medskip

We have examined this first scaling ansatz for reactive systems at equilibrium,
transiently relaxing to equilibrium, and at non-equilibrium steady-state.
It holds for $K$, $Q_+$, $Q_-$, and under some circumstances for $Q_s$.
Moreover, the exponents $\gamma$, $\zeta$, and $\delta$ are related through $\gamma/\delta = \zeta - b + 2$.
For bimolecular reaction systems, it becomes $\zeta = \gamma/\delta$, which is similar to the Family-Vicsek scaling law in surface roughening, $z = \alpha/\beta$.

\subsection{Example: Unimolecular decay}
\label{sec:uni}

As an illustration of the nonequilibrium scaling ansatz, again consider the irreversible reaction A$\stackrel{c}{\to}$B in a closed reaction volume, Fig.~\ref{fig:fig1}.
Initially, the vessel contains only A and at a sufficiently long time later, it contains only B.
This reaction is an event-modulated Poisson process with a propensity $a(\mathbf{x}\to
\mathbf{x}+\boldsymbol{\nu}) = cX_{\text{A}}(t)$ that decreases over time.
At early times, we find both the mean and fluctuations in the activity grow as the mixture, initially all reactant, becomes more chemically heterogeneous, Fig.~\ref{fig:fig1}.
But at long times, as the population of A is depleted, the mean saturates and the fluctuations are suppressed.

As in physical roughening, statistical correlations in the activity across the trajectory ensemble are the mechanism generating the onset of a new regime of behavior, Fig.~\ref{fig:fig1}(a).
Unlike physical roughening, however, the late time behavior in this reaction process is smoothing instead of saturation.
In fact, our numerical data agrees with the exact expression for the mean activity, $h_K(N,t) = N(1-e^{-c t})$, which goes as $Nct^1$ when $t\ll c^{-1}\ln 2$ and saturates at $N$ when $t \gg  t_\times = c^{-1}\ln 2$, Fig.~\ref{fig:fig1}(b).

Fluctuations in the dynamical activity from numerical simulations also agree with the exact expression, $w_K^2(N,t) = N e^{-c t}\left(1-e^{-c t}\right)$, Fig.~\ref{fig:fig1}(d).
Using $e^{-ct} = 1 - ct + \mathcal{O}(ct)^2$ for $ct\ll 1$, there is power law growth $w^2(N,t)/N^\gamma\sim (ct)^{\delta}$ with $\delta=1$.
Fluctuations grow to the value $w^2(N,t_\times) = N/4$ at the crossover time $t_\times = c^{-1}\ln 2$, after which they die out.
The scaling function is:
\begin{equation}
  \mathpzc{g}^{\textrm{A}\to\textrm{B}}_K
  = e^{-c t}(1-e^{-c t})
  \sim
  \begin{cases}
  (c t)^{\delta} & \text{if } t < t_\times\\
  \sfrac{1}{4} & \text{if } t = t_\times\\
  e^{-ct} & \text{if } t > t_\times.
  \end{cases}
\end{equation}
The correlations across the ``surface'' are apparent in $h_K$ and $w_K^2$, Fig.~\ref{fig:fig1}(a).
They are caused by the irreversibility of the reaction and the conserved number of molecules $N$.
While the nature of the correlations is different, the result is reminiscent of ballistic deposition where the interfacial width saturates because of lateral correlations that develop from finite system size and irreversible particle deposition~\cite{Barabasi1995}.

From our exact expressions and numerical data, the mean and variance of the dynamical activity $K$ are system size extensive such that $\gamma=1$.
With the molecularity $b=1$, the exponential arguments above give $\zeta=0$.
Fig.~\ref{fig:fig1} shows the initial power law growth of the scaling functions as $t^\delta$ with $\delta=1$.
With these exponents, scaling $w_K^2\to w_K^2/N^\gamma$ ($h_K\to h_K/N^\gamma$) gives data collapse for a given stochastic rate constant $c$, Figs.~\ref{fig:fig1}(c, e).
Together these results suggest $\gamma=\delta=1$, $\zeta = 0$, and the definition of the crossover time $t_\times = c^{-1}\ln 2$.
We can then conclude that the dynamic scaling ansatz in Eq.~\ref{eqn:ansatz} holds for the dynamical activity of any unimolecular, irreversible reaction, regardless of the nature of the reactant A or product B.

For this reaction, we have focused on the dynamical activity because its mean and variance as a function of time can be found analytically.
The scaling exponents follow immediately from the exact expressions for $h_k$ and $w_K^2$.
However, because there are only A$\to$B transitions, the observables $Q_+$ and $Q_s$ are less useful.
There is no branching along stochastic trajectories, only one path is possible,
and the only randomness is in the stochastic time sequence of A$\to$B events; hence, $Q_+$ is zero for all times.
Because of the irreversibility of this reaction, $Q_-$ is undefined.
The entropy flow $Q_s$ diverges because of the violation of detailed balance at the stationary state.
These observables are still of interest, however, for larger reaction mechanisms, provided the mechanism supports a stationary state with detailed balance.

With the dynamical activity, this example of unimolecular decay also illustrates another layer of universality.
When the reaction vessel is thermostatted, the stochastic rate constant in $t_c$ is a function of temperature, $\kappa=c(T)$.
Scaling time $t\to t/t_c = \kappa tN^\zeta$ by the characteristic time, $1/cN^\zeta$, gives further data collapse of the fluctuations in activity at different temperatures, Fig.~\ref{fig:fig1}~(insets).
Therefore, we can strengthen our conclusion: not only does the dynamic scaling ansatz in Eq.~\ref{eqn:ansatz} holds for any unimolecular, irreversible reaction regardless of the nature of A and B, it also holds for a reaction vessel at any temperature.

\begin{figure}[t!]
\centering
\includegraphics[width=1.0\columnwidth,angle=0,clip]{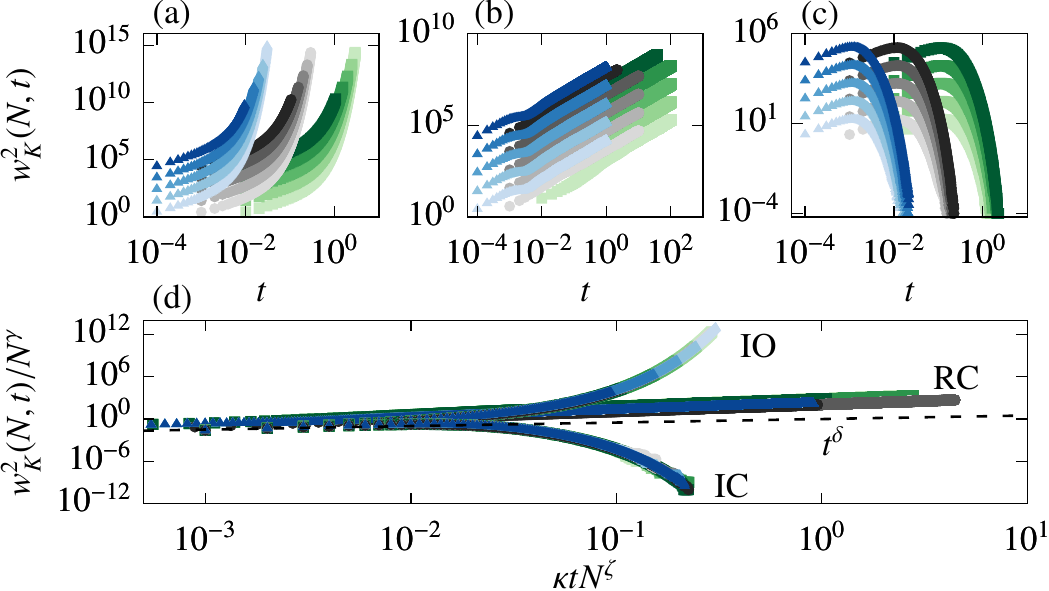}

\vspace{-0.1in}

\caption{\label{fig:fig2}\footnotesize{\textit{Effect of experimental
conditions and the reaction reversibility on the scaling function of a
bimolecular, autocatalytic reaction.} Activity fluctuations $w_K^2(N,t)$ as a function of time when the reaction is (a) irreversible in a vessel that is open $\overline{\textrm{A}}$+B$\rightarrow$2B (IO), (b) reversible in a closed reaction vessel and at dynamic equilibrium, A+B$\rightleftharpoons$2B (RC), and (c) irreversible in a closed vessel A+B$\rightarrow$2B (IC).
In all cases, $c =\sfrac{1}{10}$ (green), 1 (black), and 10 (blue), with a reverse rate constant of one for RC conditions and $N=10^2$, $10^3$, $10^4$, $10^5$, and $10^6$.
(d) Scaling fluctuations $w_K^2(N,t)/N^\gamma$ and time $t/t_c = \kappa t N^\zeta$
collapses data at all times.
The scaling function $\mathpzc{g}_K$ depends on both the reaction reversibility and molecularity and the experimental conditions, behaving as $\mathpzc{g}_K\sim t^\delta$ at all times for RC conditions and at early times otherwise.}}

\vspace{-0.1in}

\end{figure}

\subsection{Family-Vicsek form of the scaling function and analogy with kinetic roughening}

This scaling ansatz, which we will show below holds more generally, has features in common with the physical process of surface roughening where the interfacial width is used to analyze the formation and growth of the surface morphology~\cite{Barabasi1995}.
For each observable we consider, there is an analogy between chemical kinetics and physical surface roughening phenomena: each reaction along a trajectory adds to the height of the $x$th simulation column; Fig.~\ref{fig:fig1}(a) shows the surface for the cumulative reaction count.
As our notation suggests, the mean of thermodynamic observables can be considered the ``height'' of the ``surface'' and their variance as the ``interfacial roughness''.
Taking this analogy a step further for the activity, we can put the scaling relation into the form of the well-known Family-Vicsek relation~\cite{VicsekF84,*FamilyV85,DasSarma91}.
Let the initially $N$ molecules exclude a volume $L^d$ with $d$ spatial dimensions of length $L$; while we generalize here, $d=3$ in all the data we report.
We define the ``roughness'' exponent as $\alpha' := d\gamma$ and dynamic exponent as $z' := -d\zeta$ such that $z' \leq 0$; similar algebraic relationships between the $\alpha$ and $\gamma$ are known for Lyapunov exponents~\cite{PazoLP16}.
With these definitions, our ansatz takes the Family-Vicsek form:
$h_A(L,t) \sim L^{\alpha'} \mathpzc{f}_A(\kappa t/L^{z'})$ and $w^2_A(L,t) \sim L^{\alpha'} \mathpzc{g}_A(\kappa t/L^{z'})$.

An important difference with the Family-Vicsek scaling function is that the scaling functions for chemical reactions do not tend to saturate like those in physical surface roughening.
The cause of this difference is our chosen observables, the reaction mechanisms, and the characteristic timescale, all of which motivate different scaling exponents.
The scaling functions $\mathpzc{f}_A$ and $\mathpzc{g}_A$ depend on both the nature of the chemical reaction (reversibility, molecularity) and the chosen experimental reaction conditions (open, closed), Fig.~\ref{fig:fig2}.
As a consequence, the scaling law is $\alpha'/d\delta = -z'-d(b-2)$.
For bimolecular reaction systems, however, it becomes $-z'=\alpha'/d\delta$.
This form of the scaling law only differs in sign from that in surface roughening, $z= \alpha/\beta$; the sign difference is the result of the $N^{-\zeta}=L^{-d\zeta}$ dependence of our timescale $t_c$ instead of the $L^z$ dependence of the crossover time used in surface roughening.

In surface roughening, the crossover time is used to scale time and establish scaling relations.
There, the crossover time diverges as a power law and is related to the correlation length $\xi$ through $t_\times\sim L^z \sim \xi^z$ with the dynamical exponent $z$.
It follows then that the correlation length $\xi \sim t_\times^{1/z}$.
Here, we use $t_c$ to scale time in the scaling relations for well-mixed chemical reactions.
Following this reasoning suggests $t_c\sim 1/L^{d\zeta} \sim 1/\xi^{d\zeta}$.
Since $\zeta=b-1$, we can then deduce that $\xi_{\text{uni}} \sim \mathcal{O}(1)$ for unimolecular reactions (reactions are independent of the spatial extent of the volume excluded from the reaction vessel), $\xi_{\text{bi}} \sim t_c^{-1/d}$ for bimolecular reactions, and $\xi_{\text{ter}} \sim t_c^{-1/2d}$ for termolecular reactions.
For any number of spatial dimensions and a given $t_c$, the length scale $\xi$ is smaller for termolecular reactions than bimolecular reactions.
This result consistent with the intuition from collision theory that termolecular reactions tend to be less probable than bimolecular reactions because three-body collisions have a smaller cross section than two-body collisions~\cite{Gillespie92}.
Four-body collisions are so improbable that they are not typically included in chemical mechanisms.

\subsection{Homogeneous molecularity reaction mechanisms}
\label{sec:homogeneous}

Given the strength of the analogy with kinetic roughening of surfaces, we sought to test the ansatz more generally for chemical reactions other than unimolecular decay.
To start, we systematically varied the main features of the chemical mechanism and the reaction conditions, analyzing reactions both at and away from equilibrium.
For larger mechanisms, we found that the scaling ansatz holds for any set of reactions, regardless of whether the reactions are coupled in serial or in parallel or consist of cycles when the mechanism is composed of elementary reactions with the same molecularity.

\begin{SCfigure*}[][h]
\centering

\vspace{-0.1in}
\caption{\label{fig:fig3}\footnotesize{\textit{Confirmation of dynamic scaling ansatz for the thermodynamic branching observables $Q_+$, $Q_-$, and the entropy flow $Q_s = Q_+ - Q_-$ for a reaction mixture at equilibrium and relaxing to equilibrium in a closed reaction vessel.} 
The reversible reaction A$\rightleftharpoons$B occurs (a,c,e) at equilibrium and (b,d,f) relaxing to equilibrium from an initial state of all A molecules.
Mean $h_Q(N,t)\sim N^\gamma t^\delta$ of the branching observables $Q_+$, $Q_-$, and $Q_s$ as a function of time (a) at equilibrium and (b) relaxing to equilibrium.
Corresponding fluctuations $w_Q^2(N,t)\sim N^\gamma t^\delta$ as a function of time (c) at and (d) relaxing to equilibrium.
In (a-d), the forward rate constant is $c_f=0.1$ (green), 1 (black), and 10 (blue); the reverse rate constant is (a,c) $c_r=c_f$ and (b,d) $c_r=1$.
Darker colors indicate larger $N$: $N=10^2-10^6$.
(e) Sum of fluctuations $w^2_{Q_+}+ w^2_{Q_-}$ and $\operatorname{cov}(Q_+,Q_-)$ (e) at equilibrium (f) relaxing to equilibrium from an initial state of all reactant.
 In (e) and (f), $c_f=c_r=1$.
The wandering exponent is $\gamma=1$ for the mean of $Q_{+/-}$ and the variance $w^2_Q\sim N^{+1} t^{+1}$.
At equilibrium, $Q_s$ has a scaling exponent $\gamma=0$ both for its mean and variance.
During the relaxation to equilibrium $\gamma=1$ for $h_{Q_s}$; for the fluctuations, however, $w^2_{Q_s}$ has a $\gamma=1$ away from equilibrium and $\gamma=0$ near equilibrium.}}

\includegraphics[width=1.0\columnwidth,angle=0,clip]{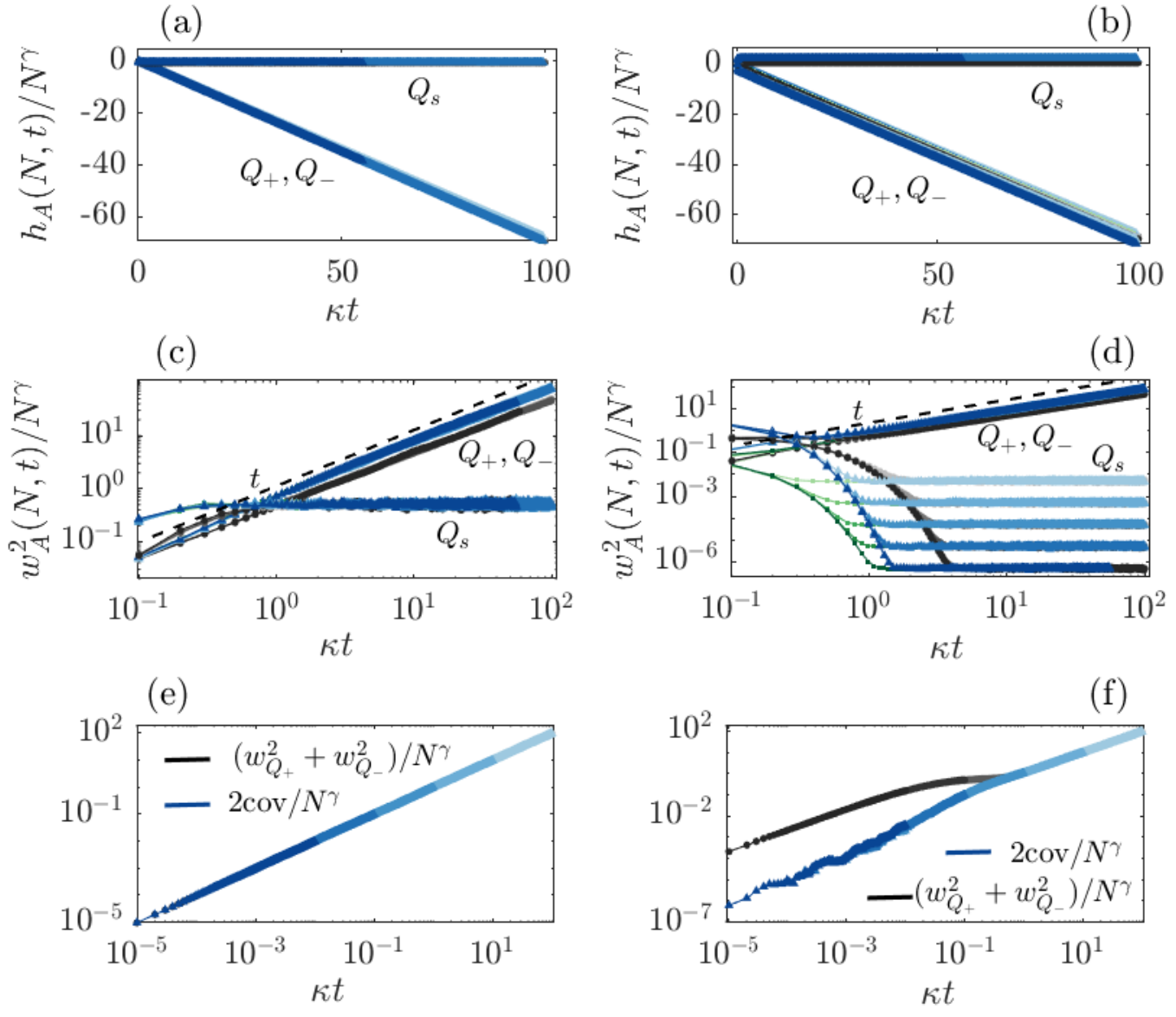}

\end{SCfigure*}

One class of reactions we considered was reversible elementary reactions at equilibrium.
Any elementary chemical reaction of the form $\sum_i \nu_WW_i +\sum_j\nu_XX_j+\ldots \leftrightharpoons \sum_k \nu_YY_k + \sum_l \nu_ZZ_l+\ldots$ obeys the scaling ansatz and has the scaling exponents $(\gamma,\delta,\zeta)=(1,1,b-1)$, Fig.~\ref{fig:fig2}.
For example, A$\leftrightharpoons$B at dynamic equilibrium has a mean and variance that agree with our scaling hypothesis for $K$, $Q_+$, and $Q_-$.
We confirmed the agreement both analytically and numerically.
As shown in Fig.~\ref{fig:fig3}, the mean and fluctuations in the branching observables diverge as $h_Q(N,t)\sim N^\gamma t^\delta$ and $w_Q^2(N,t)\sim N^\gamma t^\delta$, respectively.
Also of note is that the scaling ansatz in Eq.~\ref{eqn:ansatz} for the standard deviation give $\delta=\sfrac{1}{2}$, which agrees with growth exponent in the random deposition model~\cite{Barabasi1995}, the Gaussian universality class.

Another broad group of reactions that agree with the scaling ansatz are irreversible elementary reactions at nonequilibrium steady-state.
We considered irreversible reactions of the form $\sum_i \nu_W\overline{W}_i +\sum_j\nu_XX_j+\ldots\to \sum_k \nu_YY_k + \sum_l \nu_ZZ_l+\ldots$, where $\overline{W}$ indicates a molecular population that is constant because of an excess of reactant $W$ or permeability of the vessel walls to a reservoir of $W$~\cite{Gillespie92}.
The scaling exponents for the mean and the variance of $K$ are $(\gamma,\delta,\zeta)=(1,1,b-1)$.

Autocatalytic reactions are particularly important in combustion~\cite{newcomb2017nonequilibrium,*newcomb2018explosion} and the chemistry of living systems~\cite{Blokhuis25230}.
Well-mixed reactions of the form $X_i\to nX_i$ with branching coefficient $n$ (SM Fig.~\ref{SMfig1}) agree with the scaling ansatz.
Entire cycles of autocatalytic reactions do as well, such as the stochastic Hinshelwood cycle for cell division~\cite{IyerCSD2014} (SM Fig.~\ref{SMfig2}).
Iyer-Biswas et al.\ showed the statistics of the copy numbers and division times obey complementary scaling laws.
For an individual autocatalytic reaction in the cycle, X$\to$2X, the scaling ansatz holds for the mean $h_K(N,t)=Ne^{ct}$ and the variance in dynamical activity $w^2_K(N,t)= N e^{+c t}[e^{+c t} - 1]$, which grows as $t^{+1}$ at short times and as $e^{2ct}$ at long times, in agreement with numerical simulations.

From this survey of reactions, as we found for unimolecular decay, the scaling exponents and scaling functions depend on both the nature of the chemical reaction (reversibility, molecularity) and the chosen experimental reaction conditions (open, closed).
When density is fixed instead of volume, the scaling exponents are the same regardless of the molecularity of the reaction or the conditions $(\gamma,\delta,\zeta)=(1,1,0)$.

In our analysis of these chemistries, we also scaled the time coordinate by the parameter $\kappa$ to achieve data collapse for different choices of rate constants.
For single, reversible reaction systems that conserve the total number of molecules, we determined $\kappa$ as described in App.~\ref{app1}.
Scaling time by $\kappa$ collapses the distribution of waiting-times between reaction events; for single-reaction systems, the mean and variance of an observable $A$ for reactions with different rate constants collapse onto a single curve.
So, the system-size intensive parameter $\kappa$ is defined such that $\kappa t$ is a dimensionless time but also such that the scaling functions $\mathpzc{f}_A$ and $\mathpzc{g}_A$ are independent of the stochastic rate constants.
Figures~\ref{fig:fig2} and~\ref{fig:fig3} show that with $\kappa$, the scaling functions of $h_A$ and $w_A^2$ for different $c_f$ and $c_r$ collapse onto a single curve, independent of the nature of the chemical species and the rate constants governing the reactions.
The values of rate constants are commonly taken to be functions of temperature through an Arrhenius expression.
Within this modeling assumption, these scaling functions do not depend on the chemical species or temperature.

\subsection{Dynamic scaling of entropy}

More observables satisfy this ansatz than the data shown so far would suggest; the choice of observable representing the ``surface height'' $h_A$ is not unique.
The branching observables $Q_+$ and $Q_-$ evolve at an entropy (rate) and are related to the entropy production and flow.

\medskip
\noindent\textit{Information-theoretic entropy rates.--}For reactions where there is branching, the scaling exponents, function, and relation of $-Q_+$ and $-Q_-$ are the same as those of cumulative reaction count.
The observables $K$, $Q_+$, and $Q_-$ are extensive and, so, $(\gamma,\delta,\zeta)=(1,1,b-1)$.
For chemical reactions that are well described by Poisson processes, we can relate the scaling of $Q_+$ and $Q_-$ to another entropy, the entropy per unit time~\cite{gaspard2004time,Lecomte2007}, and determine the scaling exponents exactly.
While this entropy rate has previously been used to extract typical paths in nonequilibrium chemistry~\cite{Nicholson2016,*Nicholson2018a,*NicholsonBG2019}, its scaling has not been investigated.

As an example, take the equilibrium reaction A$\leftrightharpoons$B and assume the propensities are constant and proportional to the mean number of molecules of A and B.
In that case, we find from the thermodynamic formalism of Markov processes~\cite{Lecomte2007} that the entropy rate is $-\langle Q_+\rangle/t = h_{\text{\scriptsize{KS}}} = (a_f + a_r) \ln [(a_f + a_r)/a_f]$.
To make this result more transparent, consider
$c_f=c_r = 1$, with $a_f = X_{\text{A}}^{\text{eq}}$ and $a_r = X_{\text{B}}^{\text{eq}}$.
With these values the entropy per unit time $h_{\text{\scriptsize{KS}}} = N\ln N/X_{\text{A}} = N\ln 2$ is extensive in system size and the branching observable $\langle Q_+\rangle =-Nt\ln 2$ is extensive in system size and time.
A similar result holds for $Q_-$.
For both branching observables, the mean $(\gamma,\delta,\zeta)=(1,1,0)$.
From the thermodynamic formalism, we also find the exact scaling exponents for the fluctuations $w_Q^2(N,t) = t (a_f + a_r) \ln [(a_f + a_r)/a_f]$.
Again assuming that $c_f=c_r = 1$, the $Q_+$ and $Q_-$ fluctuations are $N t \ln (N/X_{\text{A}}) = Nt\ln 2$.
The fluctuations in the branching observable $w_Q^2/N^{\gamma}$ then grow as $t$ and $w_Q^2/N t = \ln 2$.
For both branching observables then, the mean and variance have the scaling exponents $(\gamma,\delta,\zeta)=(1,1,0)$.
These analytical predictions agree with our numerical data as shown in Fig.~\ref{fig:fig3}.

\medskip

\noindent\textit{Entropy flow.--}Another physically-relevant observable is the entropy that flows between the reaction vessel and the surroundings as the system evolves along a trajectory, $Q_s=Q_+-Q_-$, where $\langle Q_s\rangle=-\Delta S_e$.
The entropy flow for A$\leftrightharpoons$B is shown in Fig.~\ref{fig:fig3}(a,c) at equilibrium and (b,d) relaxing to equilibrium from an initial population of pure reactant A.
For all the reactions and nonequilibrium initial conditions we consider, the mean entropy flow has a $\gamma=1$ for all times, regardless of whether the system is away from equilibrium or relaxed to equilibrium, Fig.~\ref{fig:fig3}(b).
The entropy flow, however, is a cumulative quantity and reflects the path and initial conditions; when the mixture is at equilibrium for all times, the mean entropy flow is zero and scales as $N^0$, Fig.~\ref{fig:fig3}(a).

Unlike the other observables we consider, the wandering exponent $\gamma$ of the entropy flow fluctuations varies in time as reaction mixtures relax to equilibrium.
At $t=t_0$, we find good data collapse with $\gamma=1$ for $h_{Q_s}$ but also $w^2_{Q_s}(t_0)\sim N^{+1}$, Fig.~\ref{fig:fig3}(d).
The relative variance $w^2_{Q_s}/h_{Q_s}^2$ scales as $\sim N^{-1}$, so the entropy flow (and the heat, assuming local detailed balance) is strongly self-averaging.
As the system evolves from pure reactant A through successive nonequilibrium states to equilibrium proportions of A and B, the entropy flow fluctuations transition from $\gamma =1$ to $\gamma=0$, Fig.~\ref{fig:fig3}(b, d).

Fluctuations in the entropy flow (and production) exhibit an even richer scaling behavior that depends on the scaling of correlations between $Q_+$ and $Q_-$.
For the reactions above that are in detailed balance, the fluctuations $w_{Q_s}^2\sim N^0$ satisfy our scaling ansatz with $\gamma=0$, Fig.~\ref{fig:fig3}(c); they are independent of the system size $N$.
The relative variance $w^2_{Q_s}/h_{Q_s}^2$ scales as $N^{0}$, so the entropy flow (and the heat, assuming local detailed balance) is non-self-averaging.
Non-self-averaging behavior is commonplace in critical phenomena and is a signature of strong correlations.
While spatial correlations manifest at critical points, here the correlations are temporal correlations between the forward and conjugate reverse paths.

The distinct scaling behavior of $w^2_{Q_S}$ at and during the relaxation to equilibrium is the result of (positive) correlations between $Q_+$ and $Q_-$, Fig.~\ref{fig:fig3}.
Because $Q_s$ is the sum of two potentially correlated variables $Q_+$ and $-Q_-$, its variance is $w^2_{Q_s}=w^2_{Q_+} +w^2_{Q_-}-2[\langle Q_+ Q_-\rangle- \langle Q_+\rangle \langle Q_-\rangle]$.
The transition of the wandering exponent $\gamma$ from one to zero is understandable from the limiting cases.
As shown in Fig.~\ref{fig:fig3}(b,d), the reactant A is initially in excess, so early in the relaxation process the reaction is effectively the decay from pure A and reaction events are predominantly A$\to$B.
As we showed in Sec.~\ref{sec:uni}, this reaction has $(\gamma,\delta,\zeta)=(1,1,0)$.
In that case, the $Q_+$ and $Q_-$ are uncorrelated or weakly correlated, $w^2_{Q_+} +w^2_{Q_-}\gg 2[\langle Q_+ Q_-\rangle- \langle Q_+\rangle \langle Q_-\rangle]$, and the fluctuations scale as $w^2_{Q_s}\approx w^2_{Q_+}+ w^2_{Q_-}\sim N^{+1}$, Fig.~\ref{fig:fig3}(f).

By contrast, there is no net preference for forward or reverse reaction events at chemical equilibrium.
Any imbalance created by a forward (reverse) reaction event is soon rectified by a reverse (forward) reaction event;
if a forward reaction event occurs, there is a contribution of $+q$ to $Q_s$ and an increase in $X_{\text{B}}$, but this increase in $X_{\text{B}}$ also increases the reverse propensity, making a subsequent reverse reaction more likely to make a canceling contribution $-q$ to $Q_S$.
At long times then, when mixtures relax to equilibrium, there are strong correlations between forward and reverse reaction events, $w^2_{Q_+} + w^2_{Q_-} \approx 2\operatorname{cov}(Q_+,Q_-)\sim N^{+1}$ and $w^2_{Q_s}\sim N^0$, Fig.~\ref{fig:fig3}(c,e).

For all reactions and experimental conditions that we consider, we find the means $\langle Q_+\rangle$, $\langle Q_-\rangle$, and $\langle Q_s\rangle$ (away from equilibrium), the variances $w^2_{Q_+}$ and $w^2_{Q_+}$, and the covariance $\operatorname{cov}(Q_+,Q_-)$ all go as $\sim N^{+1}$ at all times.
However, the magnitude of the (positive) correlations between $Q_+$ and $Q_-$ determine the system-size dependence of the entropy flow $Q_s$ fluctuations.
During relaxation processes, as the reaction progresses towards equilibrium, these correlations increase as the nonequilibrium currents vanish and the wandering exponent $\gamma$ varies from $1$ to $0$.

\medskip

\noindent\textit{Entropy production.--}For reactions at equilibrium and nonequilibrium steady-states, the scaling behavior of the entropy production is identical to that of the entropy flow; as can be seen from Eq.~\ref{eq:enpro}, they are equal up to a sign under these conditions.
Specifically, the scaling exponents, the scaling function, and the relation of $\Delta S_e$ and $\Delta S_i$ are the same for mixtures at chemical equilibrium.
We confirmed this by treating a vessel as open reservoirs of A and B molecules, which could transform reversibly A$\rightleftharpoons$B, setting $c_f=c_r$ and the numbers of each species at time-independent nonequilibrium values, $X_{\text{A}}\neq X_{\text{B}}$.

Given that our dynamic scaling approach applies to transient phenomena, we again consider the relaxation to equilibrium of a reversible reaction, A$\rightleftharpoons$B, in a closed reaction vessel.
The vessel initially contains pure A, so $P[\mathbf{X}(t_0)]=1$.
At a time $t\gg (c_f+c_r)^{-1}$, the mixture will reach chemical equilibrium with a stationary probability distribution given by $P^{\text{eq}}(N,X_{\text{A}})= (1+c_f/c_r)^{-N} (c_f/c_r)^{X_{\text{A}}} \binom{N}{X_{\text{A}}}$~\cite{seifert_fluctuation_2004}.
Taking $c_f=c_r$, the stationary population is $X_{\text{A}}^{\text{eq}}=N/2$.
Putting these details together with Eq.~\ref{eq:enpro} gives the entropy production for the relaxation process, SM Fig.~\ref{SMfig3}.
The dependence of the logarithm of the ratio of the initial and final probabilities on $N$ is negligibly small compared to $Q_s$.
The entropy production then has the same scaling function, relation, and exponents in this case.

\medskip

To summarize our scaling theory so far, many common chemical reactions
have stochastic thermodynamic observables with a mean that scales as $h(N,t) \sim N^{\gamma} \mathpzc{f}_A(\kappa t
N^\zeta)$ and a variance that scales as $w^2(N,t) \sim N^{\gamma} \mathpzc{g}_A(\kappa t
N^\zeta)$, provided the observable is intensive or extensive in the usual sense.
Path dependent quantities, including the stochastic entropy flow and production have wandering exponents for the relative variance that vary between the equilibrium value of zero and the nonequilibrium value of one away from equilibrium, suggesting they self-average more strongly away from equilibrium that at equilibrium.
For the extensive observables, $K$, $Q_+$, and $Q_-$, the scaling exponents are unique to the chemical reaction and not dependent on time, the chemical species involved, or the reaction conditions.
The molecularity of the reactions involved, the nature of the
reaction vessel, and the structure of the reaction network are encoded on the form of the scaling functions $\mathpzc{f}$ and $\mathpzc{g}$.

To determine whether there are dynamic scaling exponents unique to equilibrium and non-equilibrium reaction conditions and strengthen the connection to the Family-Vicsek ansatz, we took another approach based on methods of time-series analysis~\cite{kantz2003}.

\section{Dynamic scaling of fluctuations in time series (Type II)}

Building on our ansatz above, we also consider the scaling of the temporal fluctuations in the time series associated with thermodynamic observables.
Here, the value of the scaling exponents contain more information about the (non)equilibrium nature of the reaction; their values indicate whether the mixture is away from equilibrium but also whether the observable of interest indicates a relaxation to equilibrium or feedback from autocatalysis.
Similar scaling approaches have been applied to the time series generated by the dynamics of other complex systems, such as stress-strain behavior associated with paper damage, radar backscattered signals from soils, daily crude oil price records, and daily stock market index~\cite{balankin2007dynamic}.
Through this alternative scaling approach, the fluctuations of many real-world
time-series follow Family-Vicsek dynamic scaling ansatz~\cite{balankin2007dynamic}.
Here, analyzing the time series associated with thermodynamic
observables (cumulative reaction count, entropy flow, and the branching observables) shows the conditions under which chemical reactions also satisfy this ansatz.

In this approach, we analyze the fluctuations of time series $A(x,N,t)$ using the log-returns, $\ln[A(t+\tau)/A(t)]$ with time interval $\tau$ as a scaling variable, $\tau_{\text{min}} \leq \tau \leq \tau_{\text{max}}$.
Log-returns are frequently used in analyzing financial time series where their absolute values can exhibit long-range power law correlations~\cite{plerou,Krawiecki}.
This transformation is common because it leads to a variable with time additivity and that approximately satisfies the raw-log equality (i.e., for short periods log-returns become approximately equal to raw returns).
Mathematically, it is similar to Lyapunov exponents, measuring the exponential rates of separation of deterministic trajectories~\cite{Pikovsky94}.

Here, we again calculate $A(t):=A(x,N,t)$ from our stochastic simulations over a long, observation time $t_f$ for each reaction of interest.
Following work in time-series analysis, we transform our observables using the absolute log-return of each observable $A(t)$ for a fixed time interval $\tau$:
\begin{equation}
  R_A(t,\tau) = \left|\ln\frac{A(t)}{A(t+\tau)}\right|. 
\end{equation}
We will analyze the scaling of these log returns for each observable $A$.
As before, we will consider a variety of reactions and experimental conditions, varying the rate constants and total number of molecules $N$.

\subsection{Global dynamic scaling ansatz}

Universal behavior of roughening surfaces is well known for a variety of models in the interfacial fluctuations around the mean height, the \textit{global} interface width.
Here, we again use an analogy to analyze the scaling properties of stochastic thermodynamic observables, taking the absolute log-return $R_A(t, \tau)$ as the moving interface.
A similar scaling approach has been used in the analysis of finite-time Lyapunov exponents (FTLEs), important measures of the sensitivity to initial conditions and characteristic of deterministic chaos.
In chaotic spatially-extended and dissipative dynamical systems, the evolution of perturbations, i.e., Lyapunov vectors have been shown to be formally equivalent to the evolution of a rough interface with fluctuations~\cite{PazoLP16b,Pikovsky94,Pikovsky1998,PazoL10}.

To make this analogy more precise, we model each reaction here as a continuous-time stochastic process $\hat{\mathbf{X}}(t)$ defined on the time interval $t\in [t_0,t_f]$.
From this process, we extract a discrete time-series for the observable $A$, $\{A(t_0), A(t_0+\tau), A(t_0+2\tau), \ldots, A(t_0+n\tau)\}$, by choosing the number of observations of uniform duration $T_\tau/\Delta t = n$.
Transforming these data to the discrete log return time-series $\{R(t_0,\tau), R(t_0+\tau,2\tau), \ldots, R(t_0+(n-1)\tau,n\tau)\}$ then gives the ``surface'' with statistical properties that have scaling behavior of interest.
We focus on the global ``width'',
\begin{equation}
  W_A(t, \tau) = \langle \overline{[R_A(t, \tau) -\overline{R_A(\tau)}]^2 }\rangle ^{1/2},
  \label{eqg1}
\end{equation}
as a measure fluctuations around the time average $\overline{R_A(\tau)} = t^{-1}\sum_{i=1}^{n} R_A(i\Delta t, \tau)\Delta t$ where $t_0\leq t=n \Delta t\leq T_\tau$.
The total duration of the log-return time series is $T_{\tau}=t_f-\tau_{\text{max}}$.
Angled brackets $\langle .\rangle$ denote an average over stochastic realizations.
Here, we set $t_0/\Delta t = 5,000$, $T_{\tau}/\Delta t = 20,000$, $n=15,000$, varied $\tau/\Delta t = 30, 40, 50, 60, 80$, and chose the end times, $t_0$ and $t_f$, such that $\Delta t$ is roughly timescale of a single reaction, the inverse of the total propensity $1/a_0 = t_c/N$.

Since the absolute log-return $R_A(t, \tau)$ corresponds to the moving interface in physical surface roughening, the time $t$ corresponds to the spatial variable and the interval $\tau$ to the time variable.
For the global width of the log-return from Eq.~(\ref{eqg1}), our dynamic scaling ansatz is:
\begin{equation}
  W_A(t,\tau) \sim (\kappa\tau)^{\beta} \mathcal{F}_A(\kappa t).
  \label{eqn:ansatz2}
\end{equation}
with the scaling function $\mathcal{F}_A \sim (\kappa t)^{-\theta}$ and two characteristic scaling exponents:

\medskip

\noindent (i) The \textit{``roughness'' exponent}, $\theta$, determines the power-law decay of the scaling function over $t$: $\mathcal{F}_A(t)$.
In the reactions we consider here, the exponent $\theta$ is $+\sfrac{1}{2}$, independent of the mixture composition, the total number of molecules, rate constants, the molecularity of the reactions involved, the nature of the reaction vessel or external reservoirs, and the structure of the reaction network.
For reactions at equilibrium, the total number of molecules and the rate constants, for example, determine the slope of the time series for $A$, and the log returns remove the effect of the slope magnitude on our analysis, $\ln[m(c,N)x(t+\tau)/m(c,N)x(t)]$.
Rather, the log returns are sensitive to the changes in the slope of a time series.
The global width measures fluctuations in the slope changes of a given time series.

\medskip

\noindent (ii) The \textit{Hurst exponent}, $\beta$, measures the rate at which autocorrelations decay with increasing time lag $\tau$~\cite{kantz2003}.
It depends sensitively on the reaction conditions and the form of the scaling function $\mathpzc{f}_A$.
Observables have $\beta=\sfrac{1}{2}$ if the absolute log-return time series is linear (e.g., a reversible, equilibrium reaction), indicating positive and negative autocorrelations decay rapidly in the time series.
A $\beta>\sfrac{1}{2}$ indicates an increase (decrease) in the observable is likely to be followed by another increase (decrease) on the timescale $\tau$.
We find these values for convex time series, typically autocatalytic reactions under open conditions.
A $\beta<\sfrac{1}{2}$ indicates an increase (decrease) in the observable is likely to be followed by a decrease (increase) on the timescale $\tau$.
We find these values for concave time series that saturate, such as the mean activity in closed reaction systems.

\medskip

As with our first scaling ansatz in Eq.~\ref{eqn:ansatz}, we examined the ansatz in Eq.~\ref{eqn:ansatz2} for reactive systems at equilibrium,
transiently relaxing to equilibrium, and at non-equilibrium steady-state.
It holds for $K$, $Q_+$, and $Q_-$ both at and away from equilibrium.
For $Q_s$, it holds away from equilibrium but not at equilibrium; at chemical equilibrium $\langle Q_s \rangle$ is zero and the log-returns are undefined.
The exponents $\theta$ and $\beta$ are not immediately related through a scaling law, instead depending on the experimental conditions and the nature of the nonequilibrium evolution of the mixture.

Approximating the global width at long times leads to a justification of the ansatz and the form of scaling function, App.~\ref{app2}.
To calculate the global width, we accumulate an ensemble of time series through stochastic simulations and calculate the log returns and the global width according to the expressions above.
However, we find comparable results if we use the mean observables directly in the log return $\ln \langle A(t+\tau)\rangle/\langle A(t)\rangle$; interchanging the logarithm and average over realizations.
We verified that the global widths from these two calculations have negligible differences (due to the arithmetic mean-geometric mean inequality) for sufficiently large $N$.
The differences we observe do not significantly affect the scaling function, scaling law, or exponents.
Since we consider cumulative quantities, such as the reaction count, the fluctuations in the time series of individual stochastic trajectories are negligible.
Because the fluctuations are small, the log return of the mean is comparable to the log return of a single realization.
With the minor differences between these two approaches, we are then effectively analyzing the scaling behavior of the log returns for $\langle K\rangle$, $\langle Q_{+}\rangle$, $\langle Q_{-}\rangle$, and $\langle Q_{s}\rangle$.

\vfill

\begin{figure}[!t]
\includegraphics[width=1.0\columnwidth,angle=0,clip]{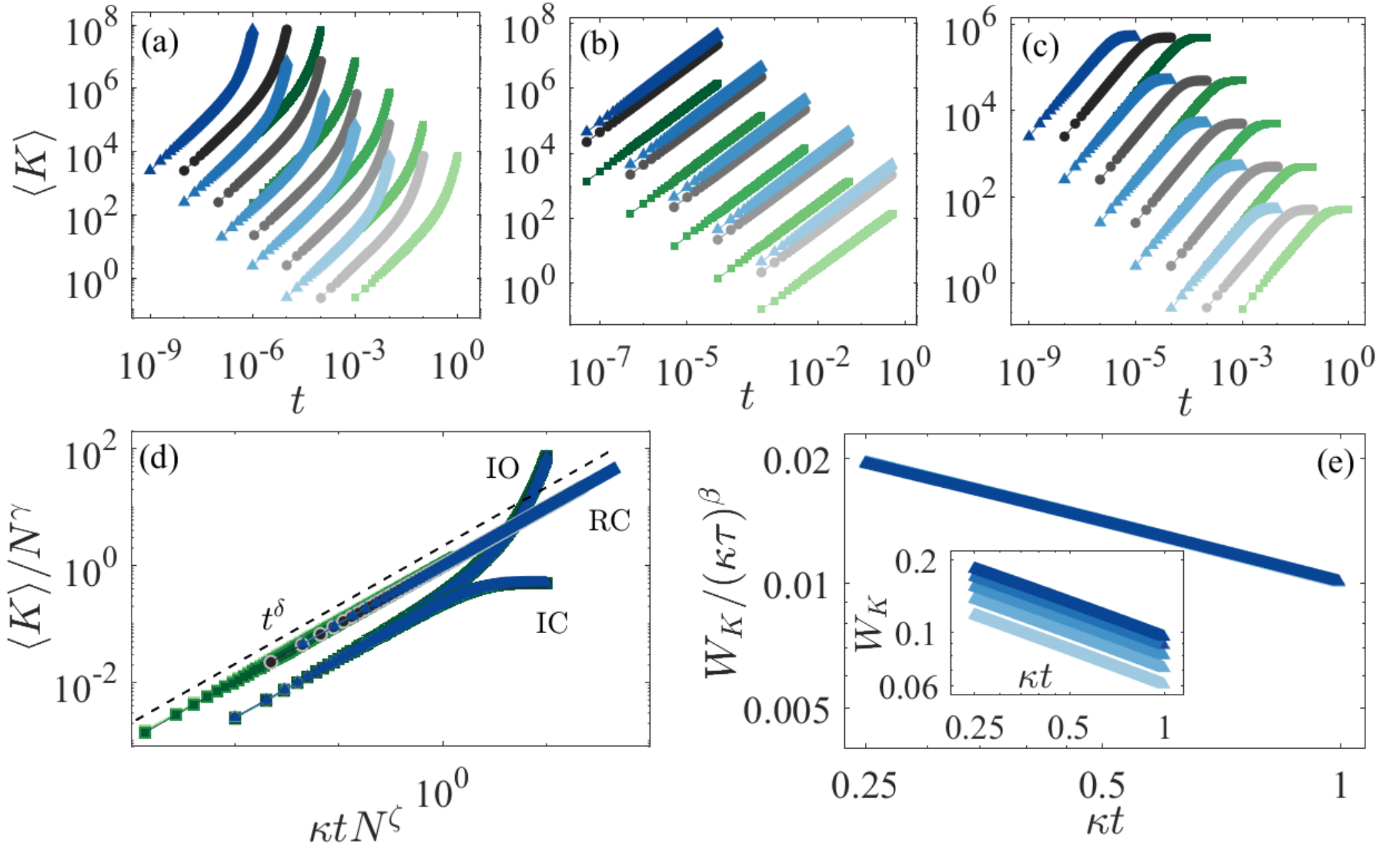}

\vspace{-0.1in}

\caption{\label{fig4} \footnotesize{\textit{Effect of experimental conditions and the reaction reversibility on the scaling function of a bimolecular, autocatalytic reaction.} Mean activity $\langle K(N,t) \rangle$ as a function of time when the reaction is (a) irreversible in an open vessel $\overline{\textrm{A}}$+B$\rightarrow$2B (IO), (b) reversible in a closed reaction vessel and at dynamic equilibrium, A+B$\rightleftharpoons$2B (RC) with $\kappa=4c_f^2c_r/(2c_f+c_r)^2$, and (c) irreversible in a closed vessel A+B$\rightarrow$2B (IC).
Colors indicate $c = c_f = \sfrac{1}{10}$ (green), 1 (black), and 10 (blue) with $c_r=1$ for RC conditions and darker colors indicating larger $N$: $N=10^2-10^6$ molecules.
(d) Scaling $\langle K(N,t) \rangle/ N^{\gamma}$ and $t/t_c = \kappa t N^\zeta$
gives good data collapse at all times.
(e) Scaling the global width $W_{K}(t, \tau)/(\kappa \tau)^{\beta}$ and time $\kappa t$ with $\kappa = c = 10$ collapse the data for the reaction that is irreversible in an open vessel $\overline{\textrm{A}}$+B$\rightarrow$2B (IO).
The inset shows the raw data.
The global width for the other reaction conditions is similar (not shown).
For (a) and (c), the mixture relaxes to a steady-state from an initial population of $X_{\text{A}}=0.99N$ and $X_{\text{B}}=N-X_{\text{A}}$.}}

\vspace{-0.1in}
\end{figure}

\subsubsection{Example: Autocatalytic reaction}

As an illustration of the ansatz in Eq.~\ref{eqn:ansatz2} and to determine how the scaling exponents depend on the reaction conditions, let us focus on a bimolecular, autocatalytic reaction. We consider the reversible reaction A+B$\leftrightharpoons$2B in a closed vessel at equilibrium and the irreversible reaction A+B$\to$2B, both when the vessel is open to a reservoir of A and when it is closed.
Fig.~\ref{fig4}(a-c) shows the mean cumulative reaction count, $K$, which complements the variance shown in Fig.~\ref{fig:fig2}.
There is good agreement with the ansatz in Eq.~\ref{eqn:ansatz} for all three cases with $\gamma=1$ and $\zeta=1$, Fig.~\ref{fig4}(d).

When the reaction is irreversible, and the vessel is open to a reservoir of A, there is good data collapse and agreement with the ansatz for the global width using $\theta=\sfrac{1}{2} $, $\beta \approx 0.53$, as shown in Fig.~\ref{fig4}(e).
The Hurst exponent $\beta>0.5$ indicates an increase in the observable, the mean activity, is likely to be followed by another increase on the timescale $\tau$, which is evident in (a).
In the reactions we surveyed, convex time series appear to be a signature of autocatalytic reactions under open conditions.
To confirm this interpretation, and the assignment of $\beta$ as the Hurst exponent, we considered the same irreversible reaction in a closed container.
There is good data collapse for the global width under these conditions with $\theta=\sfrac{1}{2}$ and $\beta\approx \sfrac{1}{4}$.
The $\beta<0.5$ indicates an increase in the observable is likely to be followed by a decrease on the timescale $\tau$, which is evident in (c).
As a reference point, we took A+B$\rightleftharpoons$2B at dynamic equilibrium in a closed reaction vessel.
This reaction is a Poisson process with good data collapse using $\theta=\sfrac{1}{2}$ and $\beta=\sfrac{1}{2}$.

Deviations from $\beta=\sfrac{1}{2}$ do occur under reversible, closed conditions, however, when the mixture evolves away from chemical equilibrium.
The global width in Fig.~\ref{fig4}(e) confirms that unlike the Type I scaling for A$\to$B, the number and values of scaling exponents giving data collapse here do not depend on the reaction molecularity.
When considering single elementary reactions, experimental conditions of the reaction vessel determine the global scaling exponents, not the form of the scaling function.
These features of the global ansatz are distinct from Type I where reaction conditions only determine the form of the scaling function, not the scaling exponents.

An important conclusion from our survey of single elementary reactions, the deviation of $\beta$ and $\theta$ from $\sfrac{1}{2}$ is a signature of the evolution of the mixture away from chemical equilibrium.
In our example, $\operatorname{sgn}(\beta-\sfrac{1}{2})$ indicates whether the mean $\langle A\rangle$ grows precipitously through a self-sustaining, autocatalytic process ($> 0$) or saturates at long times due to insufficient reactant ($< 0$).
To explore this finding further, we systematically varied the main features of the chemical mechanism and the reaction conditions, analyzing reactions both at and away from equilibrium.

\subsubsection{Homogeneous molecularity reaction mechanisms at and away from chemical equilibrium}

The bimolecular reaction above illustrates a number of features of this scaling ansatz that we find in surveying sets of reactions with the same molecularity.
For all the reactions we analyzed, the values of $\theta$ and $\beta$ are independent of the total number of molecules and the initial composition of the mixture.
Analyzing larger mechanisms, we found that this scaling ansatz also holds for any set of reactions, regardless of whether the reactions are coupled in serial or in parallel or consist of cycles when the mechanism is composed of elementary reactions with the same molecularity.

Regardless of the mechanism, in any mixture satisfying detailed balance there will be linear growth in $\langle K\rangle$, $-\langle Q_+\rangle$, and $-\langle Q_-\rangle$ with $\beta=\theta=\sfrac{1}{2}$.
Reversible elementary reactions at equilibrium of the form $\sum_i \nu_WW_i +\sum_j\nu_XX_j+\ldots \leftrightharpoons \sum_k \nu_YY_k + \sum_l \nu_ZZ_l+\ldots$ obeys the scaling ansatz and has the scaling exponents $(\theta,\beta)=(\sfrac{1}{2},\sfrac{1}{2})$.
For example, A$\leftrightharpoons$B at dynamic equilibrium has a global width that agrees with our scaling hypothesis for $K$, $Q_+$, and $Q_-$.
We confirmed the agreement both analytically, with the approximations leading to Eq.~\ref{eq:approx}, and numerically.

Detailed balance, however, is not a necessary condition for the scaling ansatz or the class with $\theta=\beta=\sfrac{1}{2}$.
We again found that the cumulative reaction count agrees with this global scaling ansatz for irreversible elementary reactions at nonequilibrium steady-state: $\sum_i \nu_WW_i +\sum_j\nu_X\overline{X}_j+\ldots\to \sum_k \nu_YY_k + \sum_l \nu_ZZ_l+\ldots$ with a reservoir or excess of $X_j$.

\begin{figure}[!t]

\includegraphics[width=0.85\columnwidth,angle=0,clip]{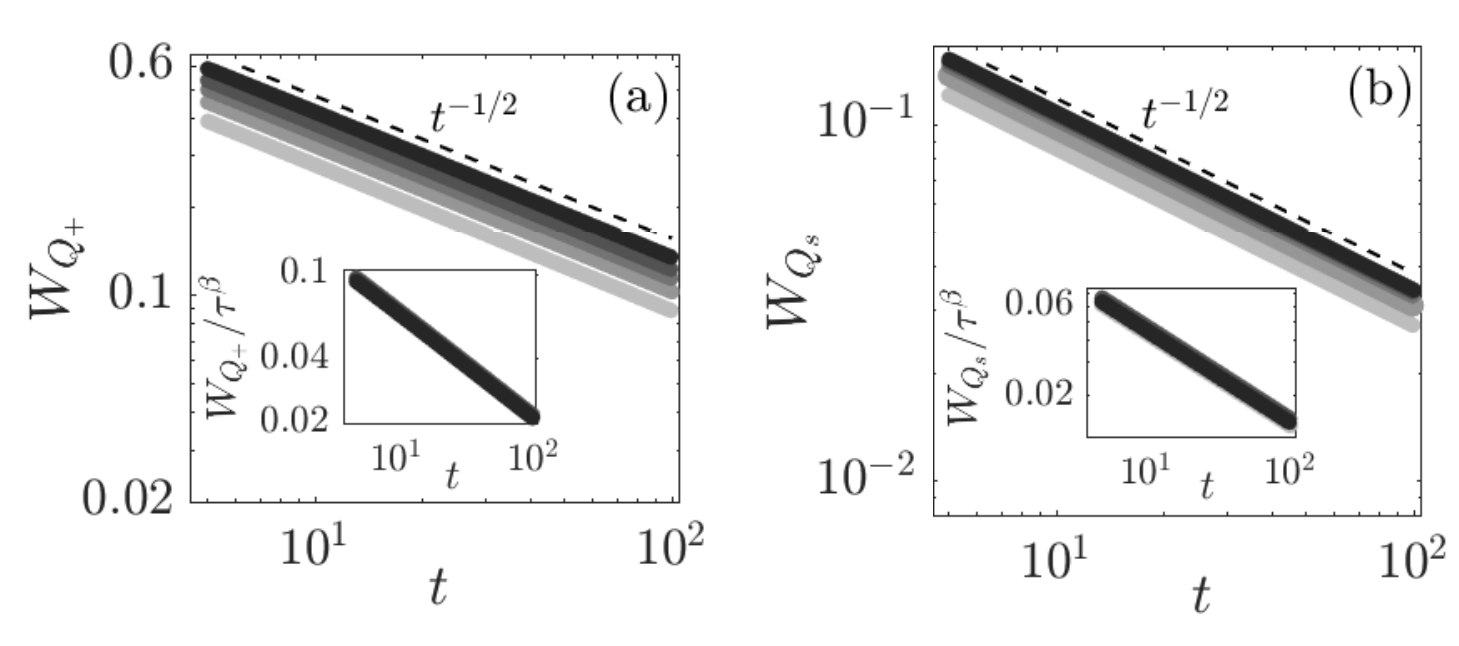}

\vspace{-0.1in}

\caption{\label{fig5}\footnotesize{\textit{Scaling of global width for the branching observables, $Q_+$ and $Q_s$ of a bimolecular autocatalytic reaction.} Rescaled global width for (a) $Q_+$ and (b) $Q_s$ for A+B$\rightleftharpoons$2B reversible reaction in a closed vessel with $c_f=1.0$ and $c_r=1.0$.
A mixture of $N=100$ molecules relaxes to a steady-state from an initial population of $X_{\text{A}}=0.99N$ and $X_{\text{B}}=N-X_{\text{A}}$.
Insets shows the global width as a function of $\tau$ and good data collapse using a $\beta$ of (a) $0.42$ and (b) $0.20$; values of $\beta<\sfrac{1}{2}$ reflect the change in sign of the slope for both $\langle Q_+\rangle$ and $\langle Q_s\rangle$ during the relaxation (SM Fig.~\ref{SMfig4}).
The original time series vary from $t=10^{-3}-10^2$, which is discretized with $\Delta t=0.001$ and $t_0/\Delta t=5000$ and $T_{\tau}/\Delta t =100,000$.}}

\vspace{-0.1in}
\end{figure}

As in Type I, the global scaling behavior for the branching observables $Q_+$, $Q_-$ is the same as that of the cumulative reaction count, Fig.~\ref{fig4}(b); others have found similar behaviors of these observables~\cite{Lecomte2007}.
And, again, the entropy flow $Q_s$ has a scaling behavior that is distinct from $Q_+$ and $K$.
When detailed balance holds, the entropy flow $\langle Q_s\rangle$ is zero and the exponents $\beta$ and $\theta$ are undefined.
But, when detailed balance is violated, the scaling function of $Q_s$ will be the same as that of $K$, $Q_{+/-}$.
The Hurst exponent $\beta\leq \sfrac{1}{2}$, however, will be less than for $K$ and $Q_{+/-}$, Fig.~\ref{fig5}.
For non-equilibrium processes at steady state, $\beta=1/2$.

By transforming observables into their log returns, the scaling exponents are dependent upon the curvature across the time series, loosely speaking.
However, because of this choice of transformation, one class of time series is not readily analyzed with this global scaling approach: reactions with observables that grow or decay exponentially $\langle A(t)\rangle \sim Ne^{\pm \kappa t}$.
For example, the simplest autocatalytic reaction, X$\stackrel{c}{\to}$2X, has the mean dynamical activity: $Ne^{ct}$.
As a result of this form of the growth, the log-returns are constant $R_A=|\ln[Ne^{ct}/Ne^{c(t+\tau)}]|=\tau$, which gives a mean $R_A$ of $\tau$ and a standard deviation of zero.
Another example, is the reaction $\overline{\text{A}}\to$B, which has a mean activity that grows exponentially.

An important reference point are linear time series, for which $\beta$ is independent of the rate constants.
As an example, the activity $K$ for Poisson processes like that shown in Fig.~\ref{fig4}(b), has a $\beta$ of $\sfrac{1}{2}$ regardless of the values of $c_f$ and $c_r$.
However, for mixtures relaxing to chemical equilibrium, the value of the $\beta$ can vary with the rate constants if they alter the concavity of the time series (or the scaling function $\mathpzc{f}$).
For example, the concave time series shown in Fig.~\ref{fig4}(c), has a $\beta$ in the range $0 \leq \beta \leq 0.5$.
Assuming an Arrhenius relation between the rate constant and temperature, this result suggests the exponent $\beta$ is temperature dependent.
This range is set by two extremes:
For reactions that are fast and complete within $\Delta t$, $\langle K \rangle$ is effectively constant and $\beta$ is zero.
For reactions that are slow compared to the duration of the time series $T_\tau$, $\langle K \rangle$ will be a roughly linear function of time and so $\beta$ will be $\sfrac{1}{2}$.
If $\langle K \rangle$ grows linearly with time and saturates before $T_\tau$, the observed value of the $\beta$ will be between $0$ and $\sfrac{1}{2}$, with the exact value depending on $T_\tau$.

These extreme cases, however, point to a way to obtain scaling exponents, $\theta$ and $\beta$, independent of the value of the rate constants.
These exponents have unique values if the final time of the time series $\kappa t_f$ and the number of data points in the discrete time series, $n=\kappa t_f/\Delta t$, are both fixed and time is scaled $t\to \kappa t$. 
We find this feature of the global scaling ansatz for any concave or convex time series (excluding exponential growth or decay).
Scaling time leads to time series for different rate constant that have the same log returns; that is, the time series are shifted vertically depending on $N^\gamma$ but otherwise identical.
As we showed in Sec.~\ref{sec:homogeneous}, scaling time by $\kappa$ can collapse $h_A/N^\gamma$ and $w^2_A/N^\gamma$ independent of the reaction mechanism, molecularity, or nature of the reactants and products.
Global scaling exponents that are independent of the values of the rate constants (and temperature for Arrhenius rates) then require the characteristic timescale $\kappa$, just as in our previous scaling ansatz.

\subsection{Local dynamic scaling ansatz}

The global width can show distinct behavior from the local interfacial fluctuations in kinetic roughening models~\cite{Ramasco2000}.
Local fluctuations are often measured by the local width or height-height correlation functions, properties that have been analyzed in surface growth models~\cite{Krug1994,Lopez1996,DasSarma1996} and experiments~\cite{Morel1998}.
Leveraging the analogy between the log returns of thermodynamic observables and the height of an abstract interface, we also test a local scaling ansatz for classes of chemical reactions.

\begin{figure}[!t]
\centering
\includegraphics[width=0.9\columnwidth,angle=0,clip]{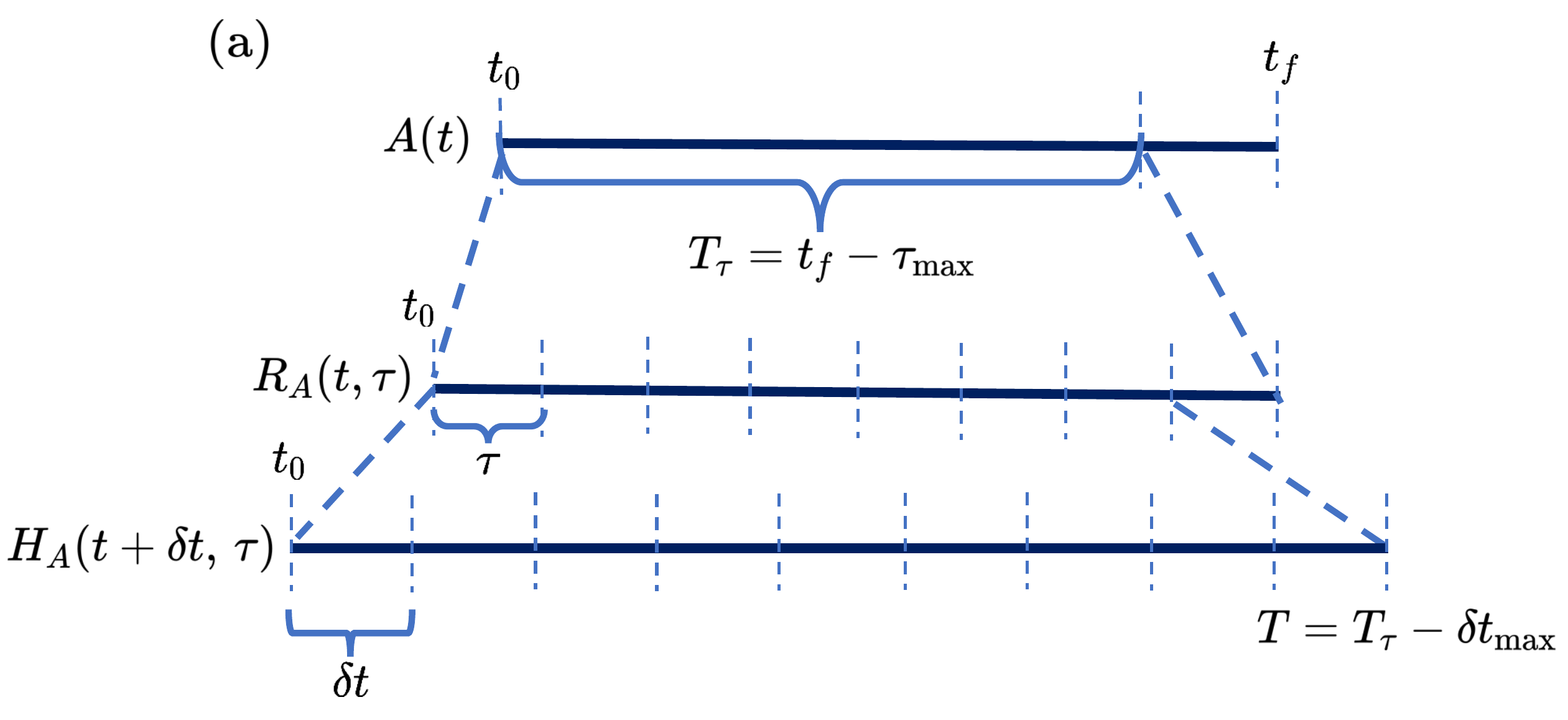}\\
\includegraphics[width=0.9\columnwidth,angle=0,clip]{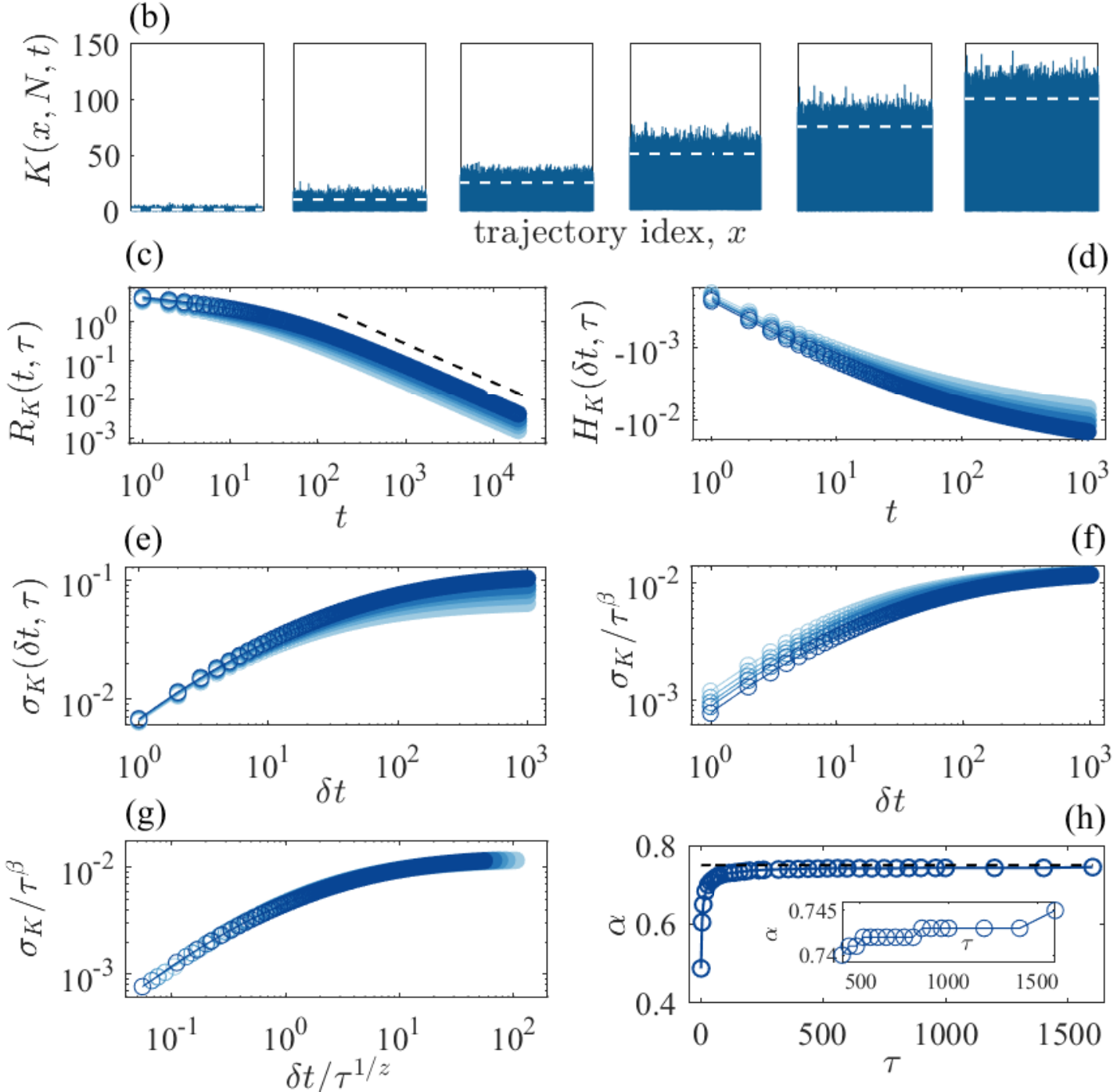}
\vspace{-0.1in}

\caption{\label{fig6}\footnotesize{Illustration of timescales and local scaling approach for A$\rightleftharpoons$B.
(a) Schematic diagram showing the discretization of a time series for an observable $A$ over the interval $[t_0,t_f]$ using the time intervals $\tau$ and $\delta t$.
(b) The cumulative reaction count across an ensemble of simulated trajectories at six different times, $t = 1, 10, 25, 50, 75, 100$. White dashed line marks $\langle K(t)\rangle$.
Here, $\Delta t=1$.
(c) Absolute log-return $|R_K|$ of the time series of length $T_\tau=19,800$, with sampling interval $\tau=30, 40, 50, 60, 80$.
Darker colors indicate larger $\tau$.
Black dashed line is the analytical expression for large $t$.
(d) The height difference $H_K(\delta t, \tau)=R_K(t+\delta t, \tau)-R_K(t, \tau)$ as a function of $t$.
The structure function (e) unscaled $\sigma_K$ and (f) scaled $\sigma_K/\tau^{\beta}$ as a function of $\delta t$.
(g) Complete data collapse after scaling $\delta t$ by $\tau^{1/z}$ with $\alpha=0.727$, $\beta=0.5$, and $1/z=0.68$.
(h) The exponent $\alpha$ converges to $0.75$ (dashed line) with increasing $\tau$.
Inset shows convergence at higher $\tau$.
Time is in $1/c$ units.}}

\vspace{-0.1in}
\end{figure}

To analyze the discrete log-return time-series and quantify correlations over a local time $\delta t$~\cite{Ramasco2000}, we define the structure function,
\begin{equation}
  \sigma_A(\delta t, \tau) := \langle {\overline{[R_A(t+\delta t, \tau) - R_A(t, \tau)]^2 }} \rangle ^{1/2},
  \label{eq3}
\end{equation}
a ``height-height'' correlation function measuring correlations over a time $\delta t$.
As in our global scaling approach, the overbar denotes the time-average over the interval $T$.
Here, $T=t_f-\delta t_{\text{max}} -\tau_{\text{max}}$, and $\delta t_{\text{max}}$  corresponds to the maximum value of $\delta t$.
We vary $\delta t / \Delta t = 1$, $2$,\ldots,$1000$.
Fig.~\ref{fig6}(a) shows the relevant times involved in scaling the autocorrelations.

The structure function scales with the sampling interval $\tau$ as $\sigma_A \sim\tau^{\beta}$.
It also exhibits power-law growth with the time interval $\delta t$, $\sigma_A \sim \delta t^{\alpha}$ up to a crossover value $\delta t_c$, beyond which it saturates, Fig.~\ref{fig6}(c).
The crossover-time scales with $\tau$ as $\delta t_c \sim \tau^{1/z}$.
Together these findings suggest the structure function satisfies the Family-Vicsek dynamic scaling relation,
\begin{equation}
  \sigma_A(\delta t,\tau) \propto \tau^{\beta} \mathcal{G}_A(\delta t/ \tau^{1/z}),
  \label{eq6}
\end{equation}
with the scaling function:
\begin{equation}
  \mathcal{G}_A \propto
  \begin{cases}
  u^{\alpha} & \text{if $u$ $<$ 1 }\\
  \text{constant} & \text{if $u$ $\gg$ 1.}
  \end{cases}
\label{eq7}
\end{equation}
As in the previous approaches, another layer of scaling is possible using the characteristic rate, $\kappa$:
\begin{equation}
  \sigma_A(\kappa\delta t, \kappa\tau) \sim (\kappa \tau)^{\beta} \mathcal{G}_A\left[(\kappa\delta t)/ (\kappa \tau)^{1/z}\right].
  \label{eq8}
\end{equation}
For a given a reaction mechanism, scaling time by $\kappa$ gives exponents that are independent of  the rate constants, provided $t_f$ and $n$ are fixed.

Three scaling exponents are needed for this local scaling ansatz:

\medskip

\noindent (i) The exponent $\alpha$ is a measure of the strength of the long-range correlations in $\delta t$ across the fluctuating time-series.
It has values consistent with the Hurst exponent associated with the time interval $\delta t$.
The well-mixed chemical systems here have $\sfrac{1}{2}\leq \alpha \leq 1$, which would indicate long-range positive autocorrelation.
For a particular $\tau$, $\mathcal{G}_A$ increases with $\delta t^\alpha$ until the crossover time $\delta t_c$.
For Poisson processes $\alpha=\sfrac{3}{4}$ and for decay processes $\sfrac{1}{2} \leq \alpha \leq \sfrac{3}{4}$.

\medskip

\noindent (ii) The \textit{Hurst exponent} $\beta$ characterizes the dependence of structure function $\sigma_A$ on $\tau$.
For example, a Poisson process $\beta=1/2$ and a decay process $\beta$ is within $0$ and $\sfrac{1}{2}$.
This exponent also appears in the global scaling ansatz, Eq.~\ref{eqn:ansatz2}.
It is the Hurst exponent measuring long-range correlations in the time-series over intervals $\tau$.

We make these assignments because the values of $\alpha$ and $\beta$ are consistent with the usual interpretation of the Hurst exponent with $\sfrac{1}{2}$ corresponding to a stochastic process with fluctuations above and below the mean being equally likely.

\medskip

\noindent (iii) The \textit{dynamic exponent} $z$ determines the value of $\delta t_c$; increasing $\delta t_c$ increases the value of $z$. For example, A$\leftrightharpoons$B has $z=\sfrac{2}{3}$ and A$\to$B has $0\leq z\leq \sfrac{2}{3}$.

\medskip

The dynamic scaling ansatz is analogous to the Family-Vicsek relation for kinetic surface roughening~\cite{Ramasco2000}, so we find that these scaling exponents satisfy the scaling law $z=\alpha/\beta$.

To explore the local scaling ansatz, we systematically varied the reaction mechanisms and the experimental conditions, analyzing reactions both at and away from equilibrium.

\subsubsection{Elementary reactions at chemical equilibrium}

To compare to our earlier scaling ans\"atze, we again consider homogeneous molecularity reaction mechanisms.
All homogeneous molecularity reactions satisfying detailed balance, regardless of the mechanism, are Poisson processes.
These reactions exhibit linear growth in  
$\langle K \rangle$, $-\langle Q_-\rangle$, $-\langle Q_+ \rangle $, $w_K^2$, $w_{Q_+}$, $w_{Q_-}^2$.
At equilibrium, $Q_s$ is zero and time independent.

Reversible elementary reactions at equilibrium of the form $\sum_i \nu_WW_i +\sum_j\nu_XX_j+\ldots \leftrightharpoons \sum_k \nu_YY_k + \sum_l \nu_ZZ_l+\ldots$ obey the local scaling ansatz.
As an example, consider the scaling hypothesis holds $K$, $Q+$, $Q_-$ for A$\rightleftharpoons$B at dynamic equilibrium, Fig.~\ref{fig6}.
Fig.~\ref{fig6}(b) shows the cumulative reaction count $K(N, t)$ across an ensemble of simulated trajectories for this reaction at equilibrium.
Fluctuations in $K$ for this class of reactions grow steadily as a function of time.
Numerical data in Fig.~\ref{fig6}(c) shows that the corresponding log returns $R_K$ 
have a change in slope around $\tau$ where $\tau/t\approx 1$.
Beyond this point, the log returns decrease as $\tau/t$ at sufficiently long times, in agreement with our expression, Eq.~\ref{eq:approxlogret}.

Keeping in mind the analogy in which the log returns are the height of the kinetically roughening surface, the structure function is composed of the ``height difference'' $H_K:=R_K(t+\delta t, \tau)-R_K(t,\tau)$, Fig.~\ref{fig6}(c,d).
Fig.~\ref{fig6}(d) shows that the height difference $H_K$ is a decreasing function of $t$.
The structure function $\sigma_K$ grows as $\sigma_K \sim \delta t^{\alpha}$ with $\alpha=\sfrac{1}{2}$ up to the crossover time $\delta t_c$, Fig.~\ref{fig6}(e).
Dividing the structure function by $\tau^{\beta}$, the time series all saturate at the same value of the ordinate $\sigma_K/\tau^{\beta}$ but at different saturation times, Fig.~\ref{fig6}(f).
Complete data collapse comes from also scaling $\delta t$ by $\tau^{1/z}$ with $z=\sfrac{2}{3}$, Fig.~\ref{fig6}(g).
The value of exponent $\alpha$, however, varies with $\tau$ and appears to approach $\sfrac{3}{4}$, Fig.~\ref{fig6}(h), suggesting $\alpha \leq \sfrac{3}{4}$.
Together these results then suggest that $\alpha=\sfrac{3}{4}, \beta=\sfrac{1}{2}$, and $z=\sfrac{2}{3}$.
We find similar results for $Q_+$, $Q_-$ to those for $K$, as shown in Fig.~\ref{fig6}.

The log returns measures changes in slope of the time series, which prevents an analysis of observables that are unchanging in time.
For example, if the reaction is at dynamic equilibrium, the mean entropy flow $Q_s$ is zero for all times.
Also, one cannot analyze the local scaling of exponentially increasing (decreasing) observables $A(x,N,t)\sim e^{\pm c t}$ because the log-returns are constant with zero slope.
One example we also considered in our global scaling was A$\to$ 2A where the mean activity grows exponentially.
If the log returns of an observable vary linearly with time, the current formulation of the ansatz local scaling gives a structure function $\sigma_A$ that is independent of $\delta t$.

\begin{figure}[!b]
\includegraphics[width=1.0\columnwidth,angle=0,clip]{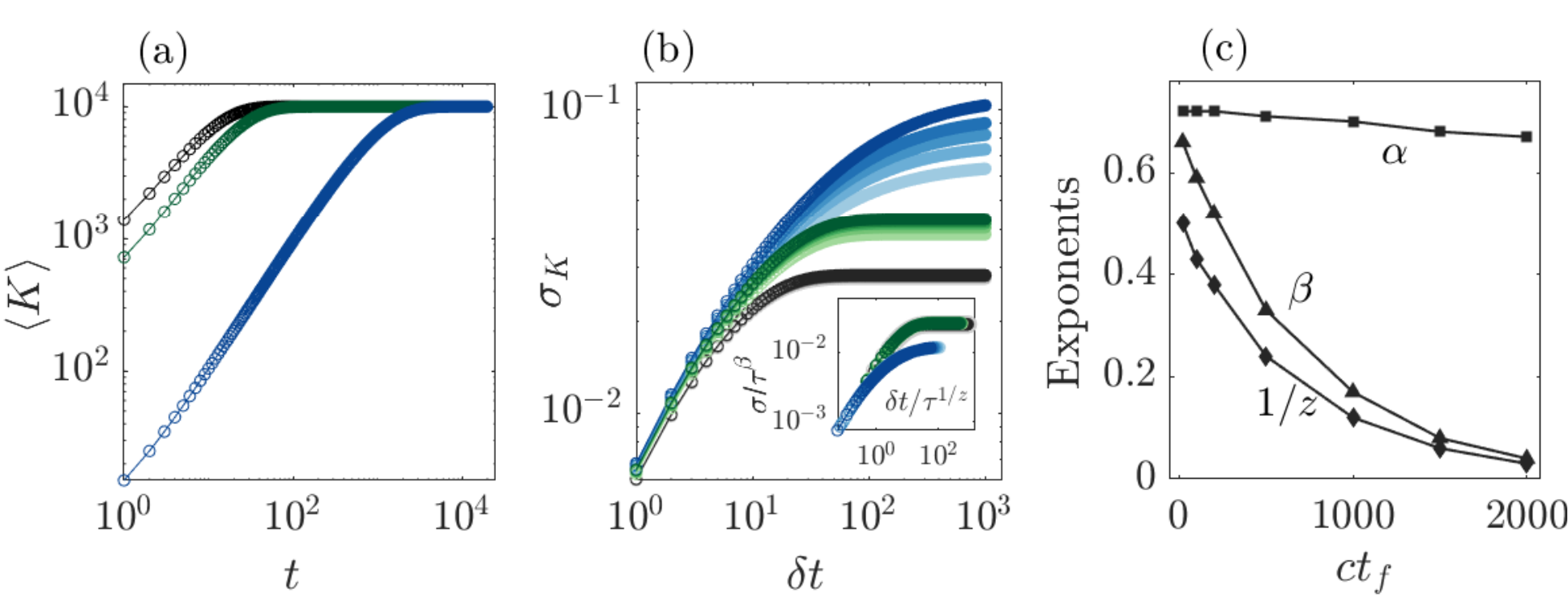}

\vspace{-0.1in}

\caption{\label{fig7}\footnotesize{Scaling of cumulative reaction count for the irreversible, unimolecular reaction A$\to$B in a closed reaction vessel.
(a) Ensemble average of the cumulative reaction count, $\langle K \rangle$ for three stochastic rate constants: $c=0.1$ (black), $0.05$ (green), $0.001$ (blue).
(b) Structure function $\sigma_K(\delta t, \tau)$ as a function of $\delta t$ with $t_f$=20,000 in $1/c$ units.
Inset shows data collapse for all $\tau$.
(c) Scaling exponents as a function of the time interval $ct_f$.
In all data shown, $X_{\text{A}}(t_0)$=10,000.}}


\vspace{-0.1in}
	
\end{figure}

While the log-returns respond to slope changes, they do not depend on the magnitude of the slope of a time series.
As a result, the log returns for a Poisson process are not determined by the rate constants or total number of molecules $\ln[(t+\tau)/t]$.
The factor $\kappa$ is not necessary for local scaling exponents that are independent of rate constants.
For Poisson processes then, if the observable is non-zero, here $Q_{+/-}$ and $K$, the local scaling exponents will be $(\alpha, \beta, z)=(\sfrac{3}{4},\sfrac{1}{2}, \sfrac{2}{3})$ independent of the number of molecules, rate constant, reaction mechanism, and the nature of the reactant and products.
For other reactions, however, such as unimolecular decay processes, $\kappa$ is necessary to uniquely determine the local scaling exponents.
With an appropriate $\kappa$, the local scaling exponents in these cases will also be independent of rate constant, number of molecules, regardless of the reaction mechanism, and nature of reactants and products; they will only depend on the duration of the time series.

\subsubsection{Irreversible elementary reactions away from chemical equilibrium}

Given that equilibrium reactions satisfy the local scaling ansatz, we consider another fundamental type of nonequilibrium reaction kinetics: irreversible unimolecular decay.
When the reaction vessel is closed and the reaction irreversible, the mean $\langle K \rangle$ does not grow indefinitely.
Instead, it grows up to a crossover time $t_\times$, after which it saturates, Fig.~\ref{fig7}(a).
Despite this difference in the mean behavior, the local scaling relation and scaling function for unimolecular decay are the same as those of reactions satisfying detailed balance.
The scaling exponents, however, only agree on short timescales where the time series is effectively linear.
On long times the unimolecular decay has $\beta < \sfrac{1}{2}$ compared to $\beta=\sfrac{1}{2}$ for equilibrium kinetics.

Local scaling exponents take on values that depend on the rate constants as well as the degree of concavity of the time series and its duration for this irreversible reaction.
Their values do not depend on the numbers of molecules.
The log-return removes their effect in this case: $R_K=\ln[N(1-e^{c(t+\tau)})/N(1-e^{ct})]$.
To determine how the scaling exponents depend on the rate constants, we considered  A$\to $B for three different rate constants at fixed $\Delta t$ and $t_f$, Fig.~\ref{fig7}(a).
We find good data collapse of the structure function $\sigma(\tau, \delta t)$ in all three cases, Fig.~\ref{fig7}(b).
However, as before, the dependence on the rate constant can be removed by scaling time by $\kappa=c$ and keeping the time interval $[\kappa t_0, \kappa t_f]$ fixed.

For a given $t_0$, because of the change in slope of $\langle K\rangle$, varying the scaled final time $ct_f$ changes the value of the exponents, Fig.~\ref{fig7}(c).
The Hurst exponent $\beta$, also in global scaling, varies from an initial value of $\sfrac{1}{2}$, when the time series is effectively linear, to zero, when the mean activity saturates because of the completion of the reaction.
Values of $\beta<\sfrac{1}{2}$ indicate an increase in $\langle K \rangle$ is likely to be followed by a decrease on the timescale $\tau$.
If $ct_f$ is such that the time dependence of $\langle K \rangle$ is linear all along the entire range, the data collapse well with the exponents characteristic of equilibrium reactions: $\alpha=\sfrac{3}{4}$, $\beta=\sfrac{1}{2}$, and $1/z=\sfrac{2}{3}$.
Increasing $ct_f$ beyond the saturation point, the mean activity deviates from linear, causing $\beta$ to decrease and, therefore, $1/z$ to decrease.
At all times, the exponents satisfy $z=\alpha/\beta$.

This example illustrates our finding that any irreversible decay process, regardless of molecularity, satisfies the local scaling ansatz with exponents 
that are independent of the numbers of molecules and rate constant.
Unlike equilibrium reactions, the breaking of detailed balance leads to more transient time series and scaling exponents with values that depend on the behavior observed.
The Family-Vicsek ansatz and the scaling law $z=\alpha/\beta$, however, still hold.

\subsubsection{Reversible elementary reactions away from chemical equilibrium}

\begin{figure}[!b]

\includegraphics[width=0.8\columnwidth,angle=0,clip]{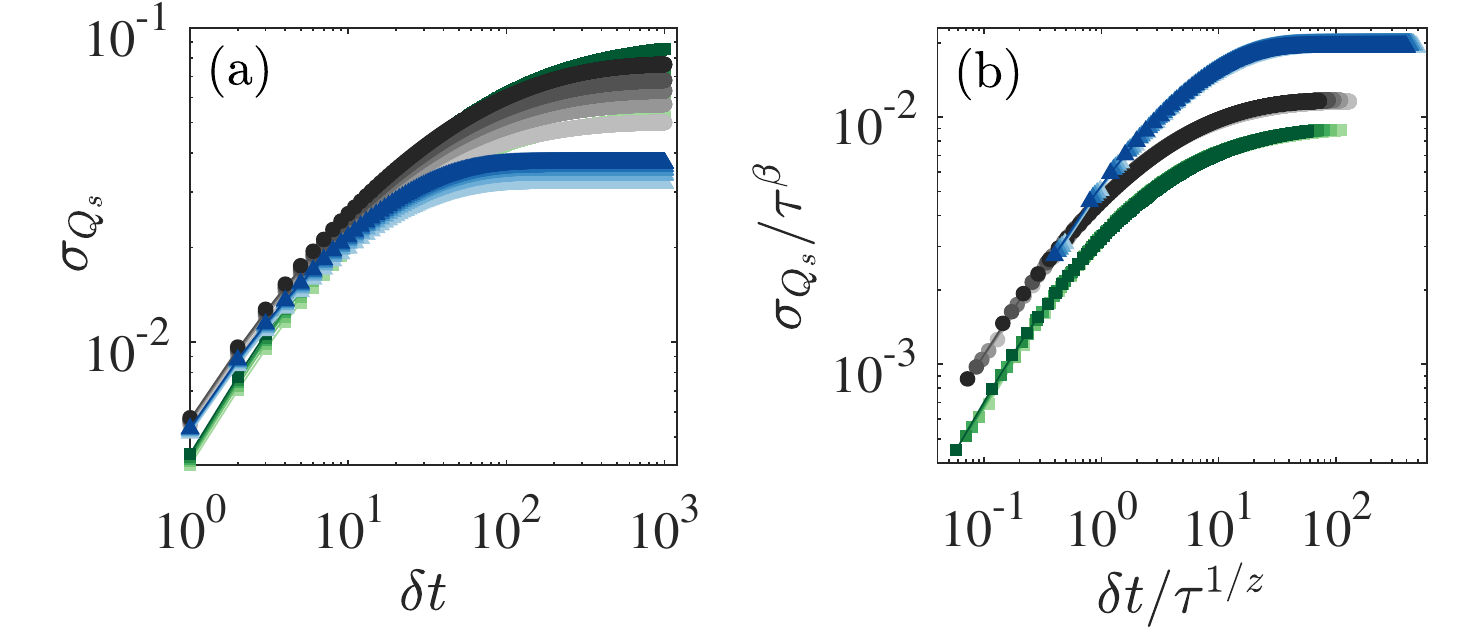}

\vspace{-0.1in}

\caption{\label{fig8}\footnotesize{Scaling of the structure function of the entropy flow for the stochastic relaxation of A$\rightleftharpoons$B from an initial population of pure reactant, A.
Initially, $X_{\text{A}}$=10,000, $X_{\text{B}}=1$.
We use $t_f$=20,000 and $c_f=c_r=c$.
(a) Structure function as a function of $\delta t$ for different rate constants $c=0.0001$ (green), $c=0.001$ (black), $c=0.01$ (blue).
(b) Data collapse well for all $\tau$ using $(\alpha,\beta,1/z)$ values of $(0.8,0.52,0.65)$ (green); $(0.71,0.43,0.6)$ (black); and $(0.69,0.15,0.21)$ (blue).}}

\vspace{-0.1in}

\end{figure}

The features of local scaling of irreversible reactions translate to reversible reactions that relax to equilibrium.
To determine how the scaling exponents away from equilibrium differ from the exponents at equilibrium for a given reaction, consider the relaxation of the reversible reaction A$\rightleftharpoons$B in a closed reaction vessel.
Initially, the reaction vessel contains pure reactant A, but at long times, the system attains equilibrium amounts of A and B.
Here, we will take the forward and reverse rate parameters to be equal without affecting our conclusions.

For any reversible reaction with homogeneous molecularity, we find that the local scaling of the activity $K$ and branching observables $Q_+$, $Q_-$ does not depend on whether the reaction is at or away from equilibrium.
These observables grow linearly on average regardless of the nonequilibrium conditions, and the scaling ansatz holds with $(\alpha,\beta,z) = (\sfrac{3}{4},\sfrac{1}{2},\sfrac{2}{3})$.

Where the picture changes is for the entropy flow $Q_s$, Fig.~\ref{fig8}.
The entropy flow $\langle Q_s \rangle$ is zero when detailed balance holds, and, as in our global scaling analysis, the local scaling exponents are undefined.
During a relaxation process, however, $\langle Q_s \rangle$ is an increasing function of time (SM Fig.~\ref{SMfig4}) and obeys the scaling ansatz in Eq.~\ref{eq6}.
The local scaling exponents are independent of molecular numbers for homogeneous reactions regardless of molecularity, again, because of the log-returns, $\ln [(h_{Q_s}(t+\tau)/N^{\delta})/(h_{Q_s}(t)/N^{\delta})]$.
The structure function and its scaling exponents depend on the rate constants for fixed $\Delta t$ and $t_f$.
However, one can identify unique exponents by scaling $t \to \kappa t$ and keeping $\kappa t_f$ and $n$ fixed.

This example is representative of the other homogeneous molecular reactions relaxing to equilibrium.
We can then expect that the dynamic scaling ansatz in Eq.~\ref{eq8} holds for the entropy flow for this class of reactions relaxing to the equilibrium at any molecular numbers, any temperature regardless of molecularity of the reactions.

\section{Complex chemical reaction}

The Type I scaling exponents found so far are unaffected by coupling reactions with the same molecularity.
Analyzing coupled reactions with two molecularities and tuning their irreversibility leads to a collision of the universality classes above.
As a representative example, we apply all three scaling approaches to an autocatalytic bimolecular reaction and a unimolecular reaction:
\begin{equation}
	\begin{aligned}
		\text{A}+\text{B} & \stackbin[c_{2}]{c_{1}}{\rightleftharpoons} 2\text{B} \\
		\text{B} & \stackbin[c_{4}]{c_{3}}{\rightleftharpoons} \text{C}.
	\end{aligned}
\label{rxn:keizer}
\end{equation}
By altering the rate constants of each elementary reaction, we can tune the (ir)reversibility of the overall reaction.
When $c_4=0$, this reaction system has stochastic and deterministic descriptions that show distinctly different behavior at steady-state -- what is known as ``Keizer's paradox''~\cite{vellela_quasistationary_2007}. While the deterministic model predicts a non-zero amount of B at steady-state, the stochastic model predicts no B in the system.
Similar autocatalytic reactions can be found in biochemical reactions, such as the phosphorylation activation of certain enzymes~\cite{Qian2010}.

In general, for Type I scaling, additional scaling exponents are necessary when chemical reactions of different molecularity are coupled.
Each molecularity has a scaled dynamic exponent $\zeta=b-1$.
Collisions and reactions between four-bodies are sufficiently improbable that they are commonly neglected in chemical kinetics.
We then only need to consider the dynamic exponent $\zeta$ for unimolecular, bimolecular, and termolecular reaction types.
However, there can be experimental conditions where reactions of a particular molecularity are dominant and a single $\zeta$ suffices.

\begin{figure}[t!]
\centering
\includegraphics[width=1.0\columnwidth,angle=0,clip]{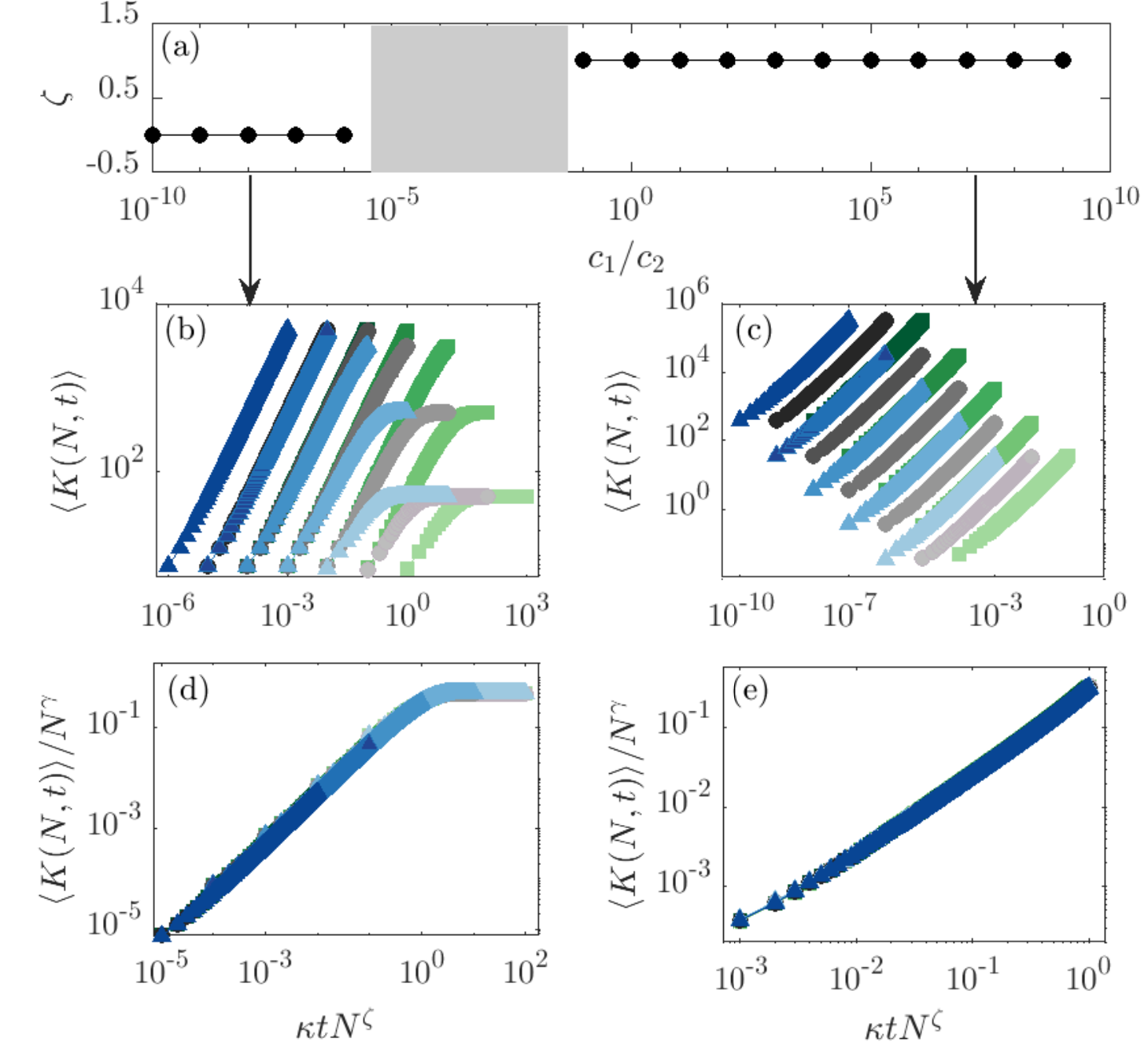}

\vspace{-0.1in}

\caption{\label{fig9}\footnotesize{Dynamic scaling (Type I) for the stochastic chemical kinetics of $\overline{\textrm{A}}$+B$\stackrel{c_1}{\rightarrow}$2B, B$\stackrel{c_2}{\rightarrow}$C when both reactions are irreversible and the vessel is open to reactant A.
(a) Variation of the dynamic exponents as a function of $c_1/c_2$ with $c_2=1$ and $N=10^2-10^6$.
(b) The unimolecular reaction dominates the mean activity as a function of time
when $c_1=1\times 10^{-9}$ and $\kappa = c_2 = 0.1$ (green), 1.0 (black), and 10.0 (blue).
Darker colors indicate larger $N$.
(c) The bimolecular reaction dominates the mean activity as a function of time when $c_2=1$ and $\kappa = c_1 = 0.1$ (green), 1.0 (black), and 10.0 (blue).
Data for all system sizes and all rate constants after scaling the mean activity $\langle K(N,t)\rangle/N^\gamma$ and time $\kappa t N^\zeta$ with (d) $\gamma=1$, $\zeta=0$ when $c_1/c_2$ is less than about
$10^{-5}$ and (e) $\gamma=1$, $\zeta=1$ when $c_1/c_2$ is greater than about
$10^{-1}$.}}

\vspace{-0.1in}

\end{figure}

As a first example, consider both reactions to be purely irreversible.
For this case, there are two Type I scaling regimes shown in Fig.~\ref{fig9}(a).
When the propensities of the first and second reaction steps are such that $a_2\gg a_1$, the unimolecular termination reaction dominates the kinetics and there is good data collapse with $\zeta=0$, Fig.~\ref{fig9}(b-d), and when $a_1\gg a_2$, the bimolecular branching reaction dominates the kinetics and there is good data collapse of $\langle K\rangle/N^\gamma$ as a function of $\kappa tN^\zeta$ with a single dynamic exponent $\zeta=1$, Fig.~\ref{fig9}(c-e).
That is, experimental conditions can exist where, despite the existence of two characteristic timescales, one timescale can dominate the kinetics and the dynamic scaling exponents.

\begin{figure}[t!]
\centering
\includegraphics[width=1.0\columnwidth,angle=0,clip]{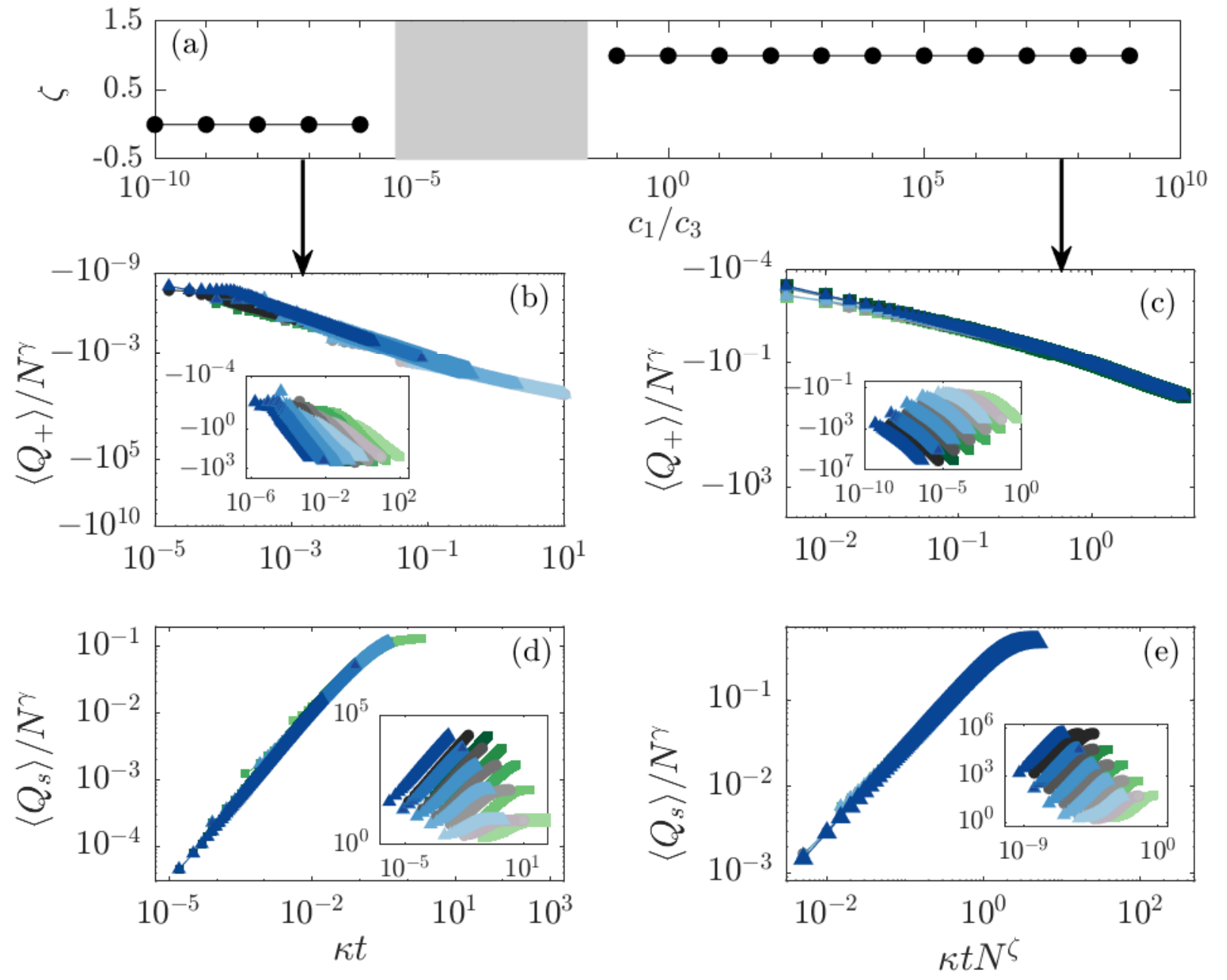}

\vspace{-0.1in}

\caption{\label{fig10}\footnotesize{%
Dynamic scaling (Type I) for the stochastic chemical kinetics A$+$B${\rightleftharpoons}$2B, B${\rightleftharpoons}$C in a closed reaction vessel.
(a) Scaled dynamic exponent $\zeta$ as a function of $c_1/c_3$ transitions from $0\to 1$, corresponding to the transition from a unimolecular- to a bimolecular-dominant reaction.
Gray region marks where a single $\zeta$ is insufficient.
Here, $c_1=c_2$ and $c_3=c_4=1$.
Data collapse for $\langle Q_+(N,t)\rangle/N^\gamma$ as a function of $\kappa t N^\zeta$ with (b) $\zeta=0$ and (c) $\zeta=1$.
Data collapse for $\langle Q_s(N,t)\rangle/N^\gamma$ as a function of $\kappa t N^\zeta$ with (d) $\zeta=0$ and (e) $\zeta=1$.
Scaling time by $\kappa$ collapses data for all rate constants.
Insets show raw data as a function of $t$.
(b,d) Unimolecular reaction events dominate the kinetics; colors indicate the value of $\kappa = c_3=c_4 = 0.1$ (green), 1.0 (black), and 10.0 (blue), $c_1=c_2=1\times 10^{-9}$. (c,e) Bimolecular reaction events dominate the kinetics; colors indicate the value of $c_1=c_2 = 0.1$ (green), 1.0 (black), and 10.0 (blue), and $c_3=c_4=0.1$.
In all panels, $N=10^2-10^6$, with darker colors indicating larger $N$.
}}

\vspace{-0.1in}

\end{figure}

Between these two regimes, the mean thermodynamic observables do not collapse using a single dynamic scaling exponent and, so, they do not satisfy the Type I scaling ansatz [marked gray in Fig.~\ref{fig9}(a)].
This regime marks a collision of two classes, here, one with $\zeta=0$ and another with $\zeta=1$.
The extent of this region depends on the range  of $N$: the range of $N$ is directly proportional to the range of rate constant ratios over which two dynamic scaling exponents are necessary.
The largest $N$ determines the $c_1/c_2$ value up to which unimolecular reaction will dominate, and the smallest $N$ will determine the $c_1/c_2$ value above which bimolecular reactions dominate.

The dominant molecularity can vary as the reaction progresses, which affect the scaling approaches we have considered.
For example, in the case where both steps in Rxn.~\ref{rxn:keizer} are irreversible and the second step is rate limiting, then the bimolecular reaction initially dominates
the scaling behavior, but at later times the unimolecular reaction dominates.
There will be a transition between the molecularity needed for the dynamic exponent $\zeta$ in the Type I scaling ansatz.
Being aware of this situation, we can identify regions where the rate parameters lead to a single dominant molecularity.
By carefully selecting the rate parameters, we can find regimes where one reaction step determines the scaling behavior for all times.

When both reactions are reversible, conditions still exist where reactions of a particular molecularity dominate the kinetics and relaxation to equilibrium.
For the reversible set of reactions, Rxn.~\ref{rxn:keizer}, the unimolecular reaction dominates the scaling when $c_1/c_2$ is less than about $10^{-5}$.
The unimolecular and bimolecular reactions compete, however, up to $c_1/c_2$ of around 0.1, where there is a regime in which the bimolecular reaction dominates the scaling.
The extent of the region where both reactions must be accounted for in the scaling is the same as that of the irreversible case, again varying with the range of system sizes.
We found similar behavior for $Q_+$ and $Q_s$, Fig.~\ref{fig10}.
Considering the form of $\kappa$ used for A$+$B$\rightleftharpoons$2B, one might expect that $\kappa$ would be a combination of all four rate constants.
However, from our analysis we found that $\kappa=c_1=c_2$ when the bimolecular reaction dominates and $\kappa=c_3=c_4$ when the unimolecular reaction dominates.
For this set of reactions, the reaction step with the highest propensity can determine the characteristic rate $\kappa$.

\begin{figure}[!t]
\includegraphics[width=1.0\columnwidth,angle=0,clip]{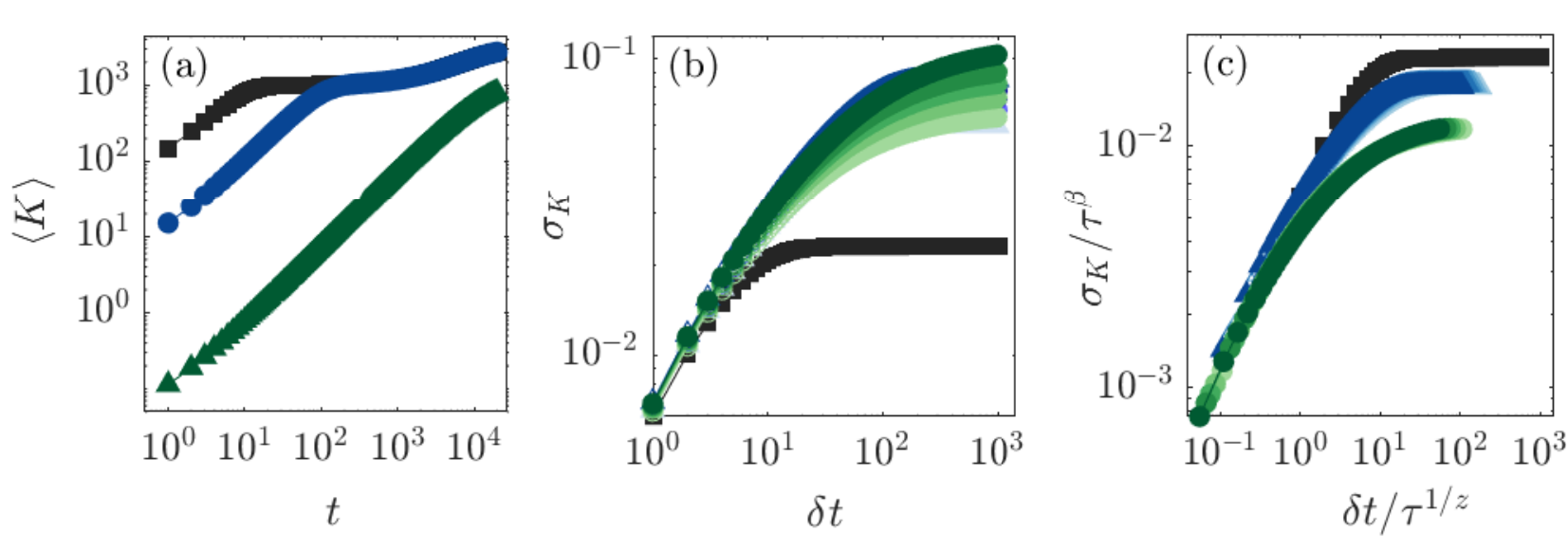}

\vspace{-0.1in}

\caption{\label{fig11}\footnotesize{Local scaling of the coupled, irreversible reactions A$+$B$\rightarrow 2$B, B$\rightarrow$C in a closed reaction vessel.
(a) The mean of the cumulative reaction count $\langle K \rangle$ when $c_2=10^{-4}$  and $c_1=10^{-4}$ (black), $10^{-5}$ (blue), $10^{-7}$ (green).
(b) Dependence of the structure function $\sigma_K$ on $\delta t$ for $\tau=30, 40, 50, 60, 80$.
(c) Data collapse of the structure function using the scaling exponents $(\alpha, \beta, 1/z)$ with values (black) $(0.66,0,0)$, (blue) $(0.73,0.36,0.49)$, and (green) $0.73,0.5,0.68$.
Initially, $X_{\text{A}}=X_{\text{B}}=1000$.}}

\vspace{-0.1in}

\end{figure}

The global and local scaling ans\"atze (Type II) are less sensitive to the dominant molecularity than the Type I scaling ansatz.
When both reactions in Rxn.~\ref{rxn:keizer} are irreversible, and the rate constants lead to a competition between molecularities, the bimolecular reaction occurs more frequently at early times, with the unimolecular reaction occuring most frequently at later times.
Fig.~\ref{fig11}(a) shows the mean cumulative reaction count $K$ for three different sets of rate constants.
Scaling the structure function $\sigma_K$ according to our local scaling ansatz gives good data collapse shown in Figs.~\ref{fig11}(b, c).
So, while the Type I ansatz holds regardless of condition, it is limited in its scope for coupled reactions of mixed molecularity.
The Type II scaling ans{\"a}tze, while their exponents are not specific to the reaction mechanism, they do hold in nonequilibrium regimes where the Type I ansatz needs refinement.

When the rate constants are identical ($c_1=c_2=1\times 10^{-4}$), the bimolecular reaction dominates initially until all reactant A has been converted into the intermediate B and unimolecular reactions begin to occur.
The local scaling exponents $\beta\approx 0$ and $1/z\approx 0$ as bimolecular reactions complete on shorter times than $\tau_{\text{min}}=30$ and time between unimolecular reactions is so large that $\langle K \rangle$ is effectively independent of time.
As a result, the log-returns are nearly $\tau$ independent and $\beta\approx 0$.
We also considered rate constants that suppress bimolecular reactions, $c_1= 1\times 10^{-8}$, in favor of unimolecular reactions, $c_2= 1\times 10^{-4}$.
In this case, the reaction observables resemble those of A$\to$B.
The mean activity, $\langle K \rangle$ grows roughly linearly as a function of time and consequently, the scaling exponents will be $(\sfrac{3}{4},\sfrac{1}{2},\sfrac{2}{3})$.
Between these two extremes, for example when $c_1=10^{-5}$ and $ c_2=10^{-4}$, the bimolecular reaction still dominates over the unimolecular reaction at early times because of the large number of A molecules and high propensity, $a_1$.
Here, in all cases, $0<\beta\leq 0.5$ indicates the initial increase in the observable is followed by a decrease or plateau on the timescale $\tau$, Fig.~\ref{fig11}.

Although they appear to be quite different, the scaling ans{\"a}tze are connected.
Because of the transition in the dominant reaction molecularity over the course of the irreversible reaction, there is not necessarily a single, unique dynamic scaling exponent $\zeta$ (Type I) for the observables we consider here.
The situation is similar in physical surface roughening where more than one dynamic exponent is necessary when the surface grows at different rates at different times or in different spatial directions.
The local scaling exponents $\alpha$, $\beta$, and $z$ also depend on the rate constants and number of molecules.
That is, while the Type II scaling apply to the time series of an arbitrary observable and reaction mechanism, data collapse through Type I is necessary for Type II exponents that are independent of the number of molecules and rate constants.

\begin{figure}[!t]
\includegraphics[width=1.0\columnwidth,angle=0,clip]{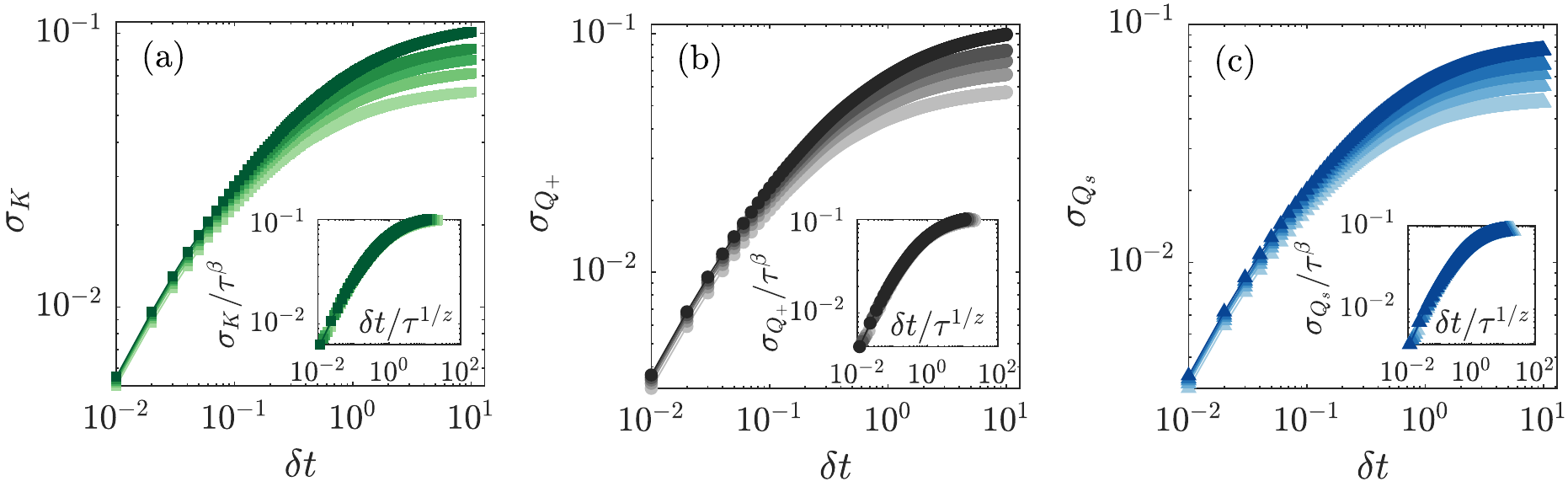}

\vspace{-0.1in}

\caption{\label{fig12}\footnotesize{%
\textit{Local scaling of three observables for the coupled reactions A+B$\rightleftharpoons2$B, B$\rightleftharpoons$C relaxing to equilibrium in a closed reaction vessel.} Structure function as a function of $\delta t$ for the (a) mean of cumulative reaction count, $K$, and (b) the branching observable, $Q_+$.
There is good data collapse for both observables using $\alpha=0.73$, $\beta=0.5$,$1/z=0.68$ (insets).
(c) Structure function as a function of $\delta t$ for the entropy flow, $Q_s$.
There is good data collapse using $\alpha=0.79$, $\beta=0.41$, $1/z=0.51$ (inset).
The mixture relaxes to equilibrium.
Initially, $N=1000$, $X_{\text{A}}= 0.75 N$, $X_{\text{B}}=N - X_{\text{A}}$, and $X_{\text{C}}=0$.
In all cases, $c_1=c_2=10^{-4}$, $c_3=c_4=10^{-1}$, and $\tau$=30, 40, 50, 60, 80.
}}


\vspace{-0.1in}

\end{figure}

Removing the perfect irreversibility, however, Type II scaling exponents are independent of the rate constants and molecular numbers for $K$, $Q_+$, and $Q_-$, regardless of whether the mixture is at or relaxing to equilibrium, Figs.~\ref{fig12}(a,b).
It is unnecessary to scale time by $\kappa$ because the means, $\langle K \rangle $ and $\langle Q_{+/-}\rangle$, are linear functions of time.
The scaling exponents are $(\sfrac{3}{4},\sfrac{1}{2},\sfrac{2}{3})$.
Again though, the situation is different for $Q_s$.
Its mean varies nonlinearly and to identify rate constant-independent scaling exponents, it is necessary to scale time $t$ by $\kappa$ and fix the final time $\kappa t_f$ and number of data points $n$ in the calculation of the structure function $\sigma_{Q_s}$, Fig.~\ref{fig12}(c).

\section{Conclusions}

Universal behaviors have been extensively explored for physical phenomena, and here we have shown that universal dynamical scaling extends to the thermodynamic observables of chemical phenomena at and away from equilibrium.
These observables satisfy three interconnected dynamic scaling ans{\"a}tze that we have tested for broad classes of chemistry from simple, elementary reactions to complex, coupled autocatalytic reactions.
Dynamical universality classes are typically determined by the dimensionality, conservation laws, symmetry of the order parameter, range of the interactions, and the coupling of the order parameter to conserved quantities~\cite{Odor2004}.
Here we find classes of well-mixed chemical reactions, do not depend on the identities of the chemical species or, in some cases, the temperature fixed by an external bath.
Instead, they are determined by the reaction vessel dimensionality, whether the vessel is open or closed, observable extensivity, and reaction molecularity.
The ans\"atze we use to define these classes include scaling laws and relations, some unknown and some known (Family-Vicsek).

The entropy flow (and production) has scaling exponents that are different at and away from equilibrium, because of the correlations between the branching observables over forward and reverse paths that emerge as mixtures approach equilibrium.
Coupling reactions capable of chemical feedback, creating collections of reactions with multiple molecularities, leads to a richer collection of scaling exponents.
In these cases, fluctuations in thermodynamic observables can transition between these classes with the variation of experimental parameters, such as temperature.
We find that despite this behavior, there are conditions where a given molecularity reaction dominates, leading to a single scaling law and universality class.
From the perspective of nonequilibrium statistical mechanics, the large catalog of chemical reaction mechanisms, while incredibly diverse, does contain universal signatures in their kinetics both at and away from chemical equilibrium.

\begin{appendix}

\section{Characteristic rate, $\kappa$, for detailed balanced systems}
\label{app1}

For single, reversible reaction systems that conserve the total number of molecules, we determined $\kappa$ using the condition of detailed balance.
As an example, take the reversible reaction, 
\begin{equation*}
  \text{A} \overset{c_f}{\underset{c_r}\rightleftharpoons}\text{B}.
\end{equation*}
The reaction is detail balanced when the forward and reverse propensities are equal: $c_fX_{\text{A}}^{\text{eq}} = c_rX_{\text{B}}^{\text{eq}}$.
When the reaction volume is closed, the total number of molecules is conserved $N=X_{\text{A}}+X_{\text{B}}$.
Combined with the detailed balance condition this constraint leads to the equilibrium propensities:
\begin{equation}
  \displaystyle\frac{a_f^{\text{eq}}}{N} = c_fX_{\text{A}}^{\text{eq}} = \displaystyle\frac{c_fc_r}{c_f+c_r}
   = c_rX_{\text{B}}^{\text{eq}} = \displaystyle\frac{a_r^{\text{eq}}}{N}. 
\end{equation}
The total propensity per molecule is $N^{-1}(a_f^{\text{eq}} + a_r^{\text{eq}}) = t_c^{-1} = \kappa N^\zeta$
and gives
\begin{equation}
  \kappa = \frac{2 c_fc_r}{c_f+c_r}.
\end{equation}
Through this approach $\kappa$ is determined for reactions where detailed balance is satisfied at equilibrium and where we can express the number of each chemical species in terms of $N$;
even for bimolecular reactions A+B$\leftrightharpoons$ C this approach does not yield $\kappa$.

\section{Justification for the global scaling ansatz}
\label{app2}

We can build upon our first scaling ansatz to justify our global scaling ansatz.
By our first scaling ansatz the log return becomes: 
\begin{eqnarray}\notag
  \ln\frac{\langle A(N,t+\tau)\rangle}{\langle A(N,t)\rangle}
  &\stackrel{\text{Eq.~(\ref{eqn:ansatz})}}{\sim}& \ln \displaystyle\frac{N^\gamma \mathpzc{f}_A[\kappa(t+\tau)N^\zeta]}{N^\gamma \mathpzc{f}_A(\kappa t N^\zeta)}\\\notag
  &\stackrel{\mathpzc{f}\sim (\kappa t N^\zeta)^\delta}{\sim}& \ln \left[\frac{\kappa(t+\tau)N^\zeta}{\kappa tN^\zeta}\right]^{\delta}\\
  &\stackrel{t\gg\tau}{\approx}& \delta\frac{\tau}{t} = R_A(t,\tau).
  \label{eq:approxlogret}
\end{eqnarray}
Our assumption in the second line is true for Poisson processes and decay processes before the mean saturates at a limiting value.
The time-average log return,
\begin{eqnarray}\notag
\overline{R_A(\tau)} 
  &=& \delta\tau t^{-1}\ln{t/t_0}
  \stackrel{t\gg\tau}{\approx} \delta\tau (t_0^{-1}-t^{-1})
\end{eqnarray}
leads to the squared width $W_A^2(t,\tau)\sim \overline{R_A^2(t,\tau)}\sim R_A(t,\tau)\tau/t_0- R_A^2(t,\tau)$.
The width squared must be non-negative, so assuming the first term dominates, we get:
\begin{equation}
  W_A^2(t,\tau)\sim \tau^{+2} t^{-1}t_0^{-1}  
  \label{eq:approx}
\end{equation}
Considering sufficiently large $t\gg \tau$, this final expression agrees with our numerical data. 
The agreement with the long time behavior justifies our dynamic global scaling ansatz.
However, because short time contributions dominate the scaling (SM Fig.~ \ref{SMfig5}), the values of the exponents differ from those we observe numerically, $\beta=\sfrac{1}{2}$ and $\theta=\sfrac{1}{2}$.

\end{appendix}

\begin{acknowledgments}

The authors acknowledge helpful conversations with Lucas~B.~Newcomb and the use of the supercomputing facilities managed by the Research Computing Group at the University of Massachusetts Boston as well as the University of Massachusetts Green High Performance Computing Cluster.

This publication was made possible, in part, through the support of a grant from the John Templeton Foundation.
It is also based, in part, upon work supported by the National Science Foundation under Grant No.\ 1856250 and the U.S.\ Army Research Laboratory and the U.S.\ Army Research Office under grant number W911NF-14-1-0359.

\end{acknowledgments}


%

\newpage 
\section*{Supplementary Material}
\begin{suppfigure}[htb!]
	\centering
	\includegraphics[width=1.0\columnwidth,angle=0,clip]{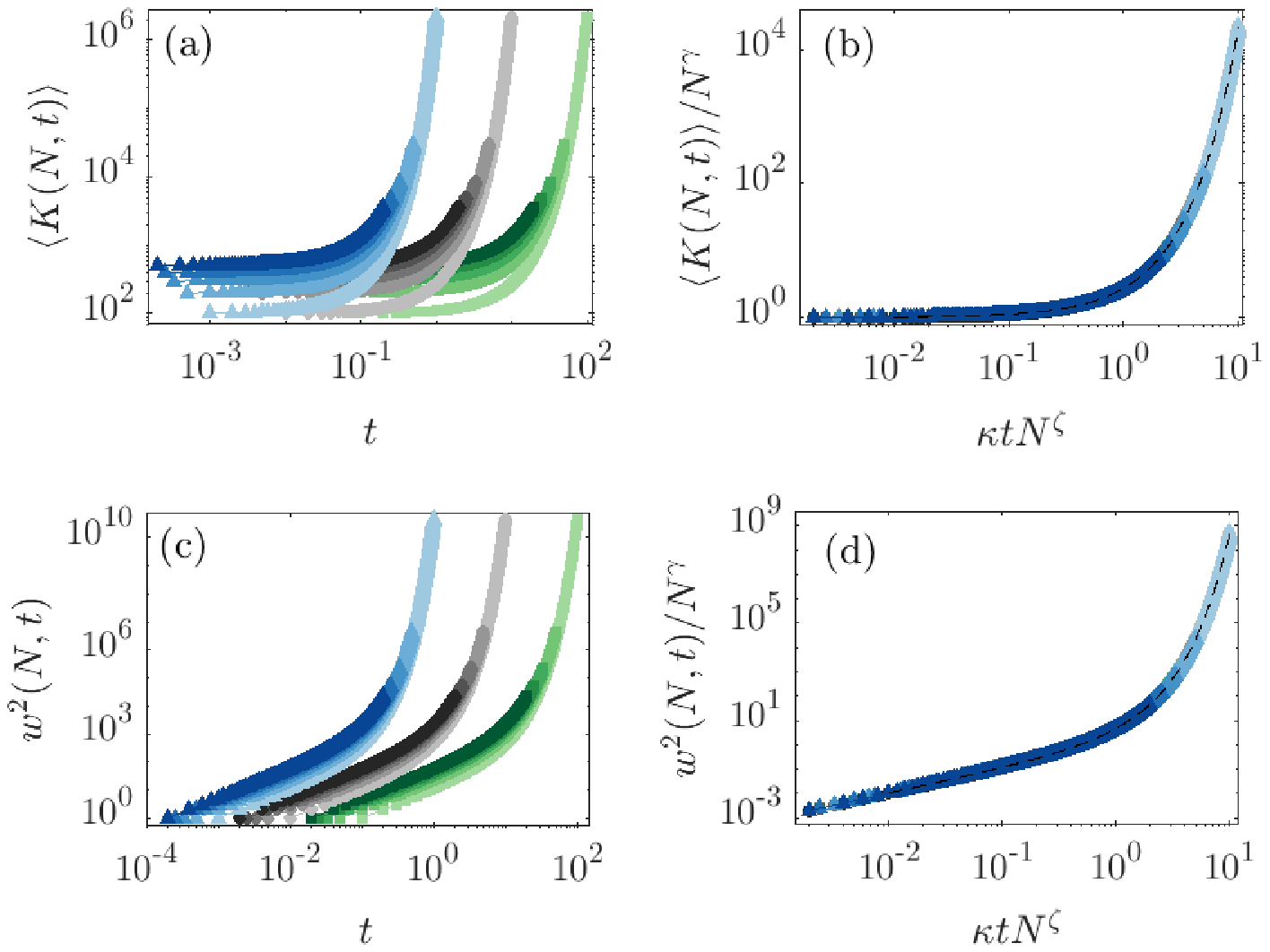}
	
	\caption{Dynamic scaling for the stochastic chemical kinetics of autocatalytic reactions $X_i\to 2X_i$.
		(a) Mean activity as a function of time for $\kappa = c = 0.1$ (green), 1.0 (black), and 10.0 (blue) and $N=100$, $200$, $300$, $400$, and $500$ molecules.
		(b) The scaled $\langle K(N,t)\rangle /N^\gamma$ with $\gamma=1$ collapses data for all system sizes onto a single curve for a given $c$.
		Scaling time by $c$ collapses data for all rate constants.
		(c) Variance of activity as a function of time for $\kappa = c = 0.1$ (green), 1.0 (black), and 10.0 (blue) and $N=100$, $200$, $300$, $400$, and $500$ molecules.
		(d) The scaled $w^2 /N^\gamma$ with $\gamma=1$ collapses data for all system sizes onto a single curve for a given $c$.
		Scaling time by $\kappa=c$ collapses data for all rate constants.}
	\label{SMfig1}
\end{suppfigure}

\begin{suppfigure}[htb!]
	\vspace{-0.1in}
	\centering
	\includegraphics[width=1.0\columnwidth,angle=0,clip]{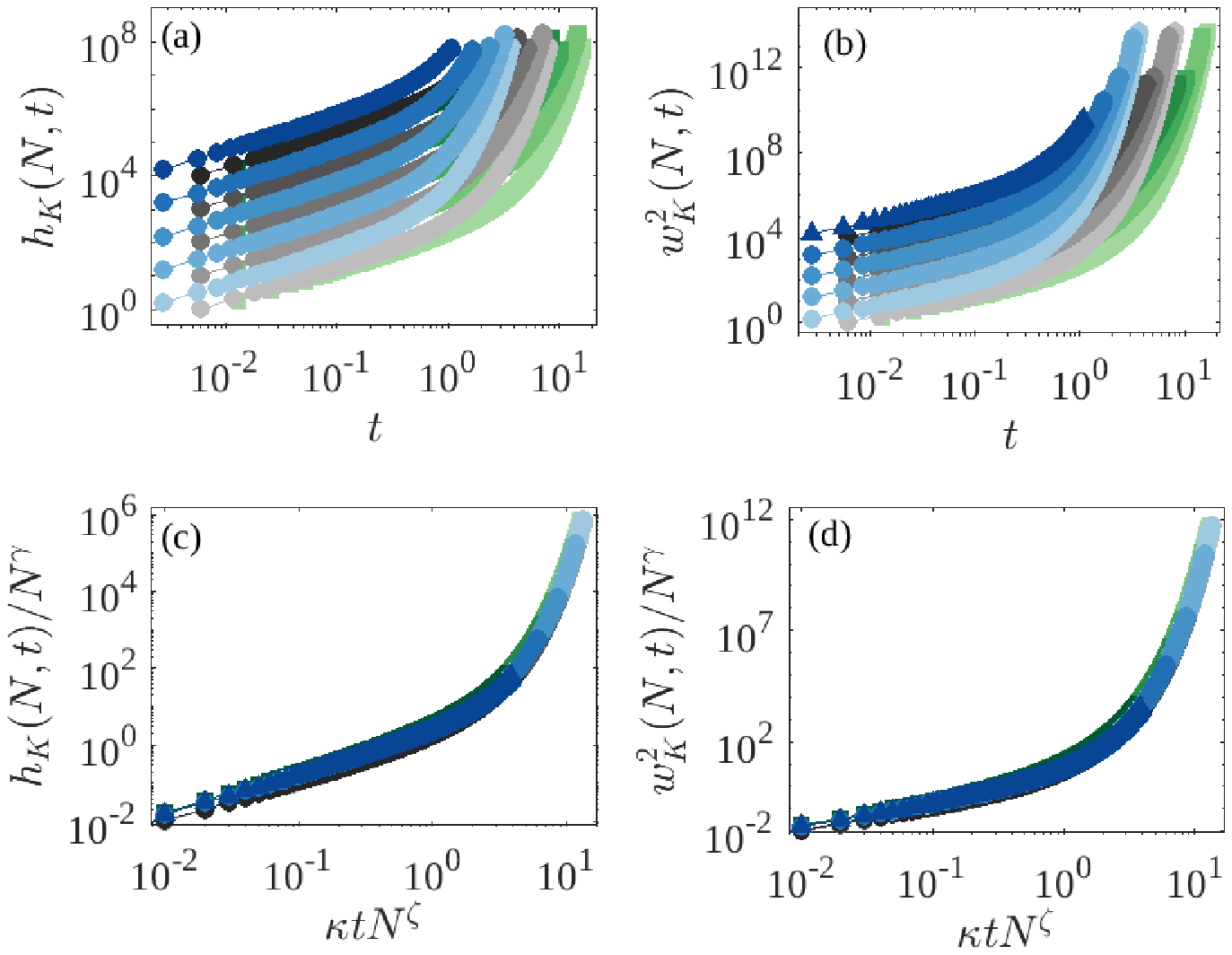}
	
	\caption{Dynamic scaling for the stochastic chemical kinetics of Stochastic Hinshelwood cycle (a) Mean activity as a function of time for $\kappa = c = 0.1$ (green), 1.0 (black), and 10.0 (blue) and $N=10^2$, $10^3$, $10^4$, $10^5$, and $10^6$ molecules.
		(b) The scaled $\langle K(N,t)\rangle /N^\gamma$ with $\gamma=1$ collapses data for all system sizes onto a single curve for a given $c$.
		Scaling time by $c$ collapses data for all rate constants.
		(c) Variance of activity as a function of time for $\kappa = c = 0.1$ (green), 1.0 (black), and 10.0 (blue) and $N=10^2-10^6$ molecules.
		(d) The scaled $w^2 /N^\gamma$ with $\gamma=1$ collapses data for all system sizes onto a single curve for a given $c$.
		Scaling time by $\kappa=c$ collapses data for all rate constants.}
	
	\label{SMfig2}
\end{suppfigure}

\newpage

\begin{suppfigure}[H]
	\centering
	\includegraphics[width=1.0\columnwidth,angle=0,clip]{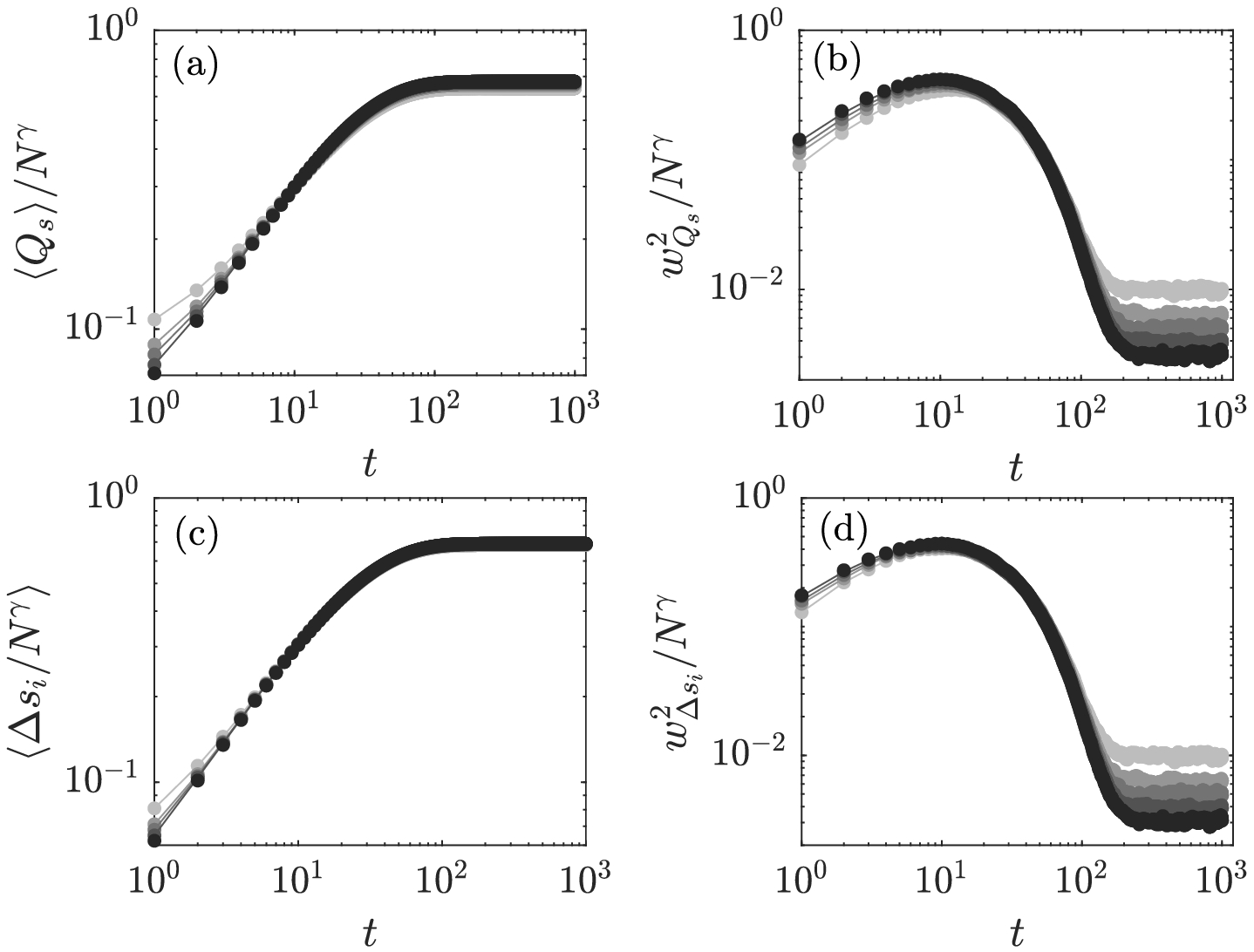}
	
	\caption{Confirmation of dynamic scaling ansatz for the thermodynamic observables  entropy flow and entropy production for a reaction mixture relaxing to equilibrium from an initial state of all A molecules. (a) Scaled mean $\langle Q_s\rangle /N^{\gamma}$ and (b) scaled variance $w^2_{Q_s}/N^{\gamma}$ as a function of time. (c) Scaled mean $\langle \Delta s_i \rangle/N^{\gamma}$ and (b) scaled variance $w^2_{\Delta_{s_i}}/N^{\gamma}$ as a function of time. The reaction vessel is closed, $c_r = c_f=0.001$, and darker colors indicate larger $N$: $N = 50, 80, 100, 130$, and $160$.
		The scaling exponent $\gamma= 1$ for all cases.}
	
	\label{SMfig3}
\end{suppfigure}

\begin{suppfigure}[H]
	\centering
	\includegraphics[width=1.0\columnwidth,angle=0,clip]{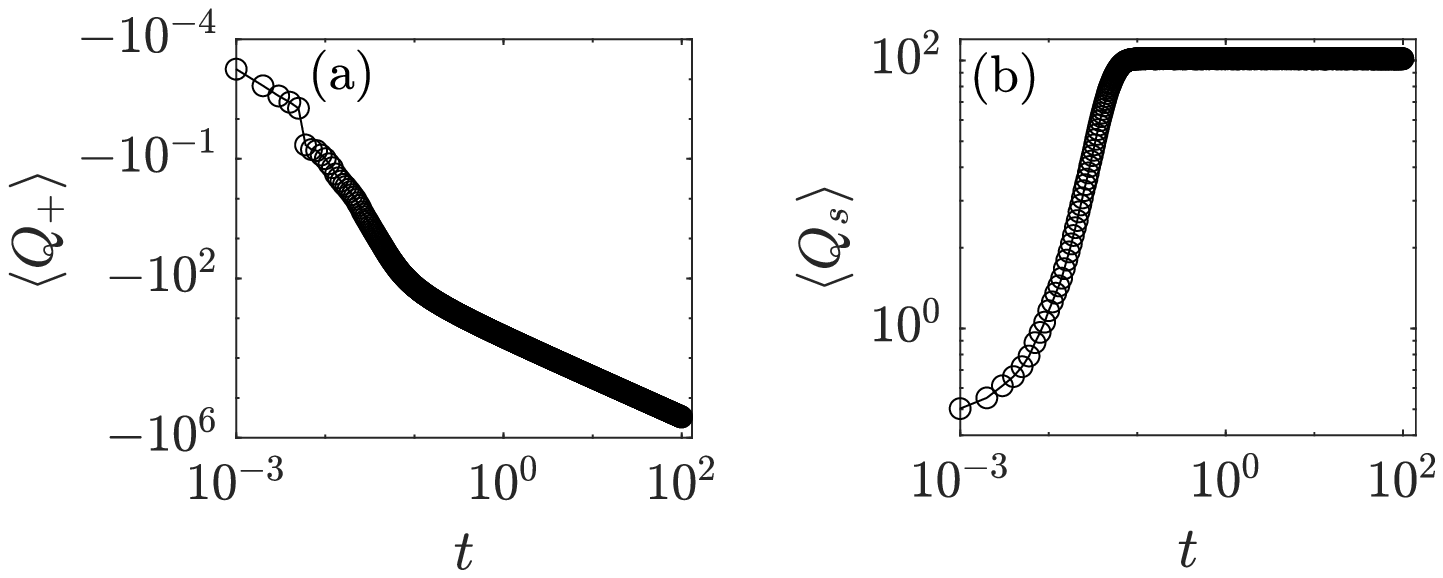}
	
	\caption{Time series for (a) $\langle Q_+\rangle$ and (b) $\langle Q_s\rangle$ for the reversible reaction A$+$B$\rightleftharpoons$2B in a closed vessel with $c_f = 1.0$ and $c_r = 1.0$.
		A mixture of $N = 100$ molecules relaxes to a steady-state from an initial population of $X_{\text{A}} = 0.99N$ and $X_{\text{B}} = N - X_{\text{A}}$.}
	
	\label{SMfig4}
\end{suppfigure}

\begin{suppfigure}[H]
	\vspace{0.25in}
	\centering
	\includegraphics[width=1.0\columnwidth,angle=0,clip]{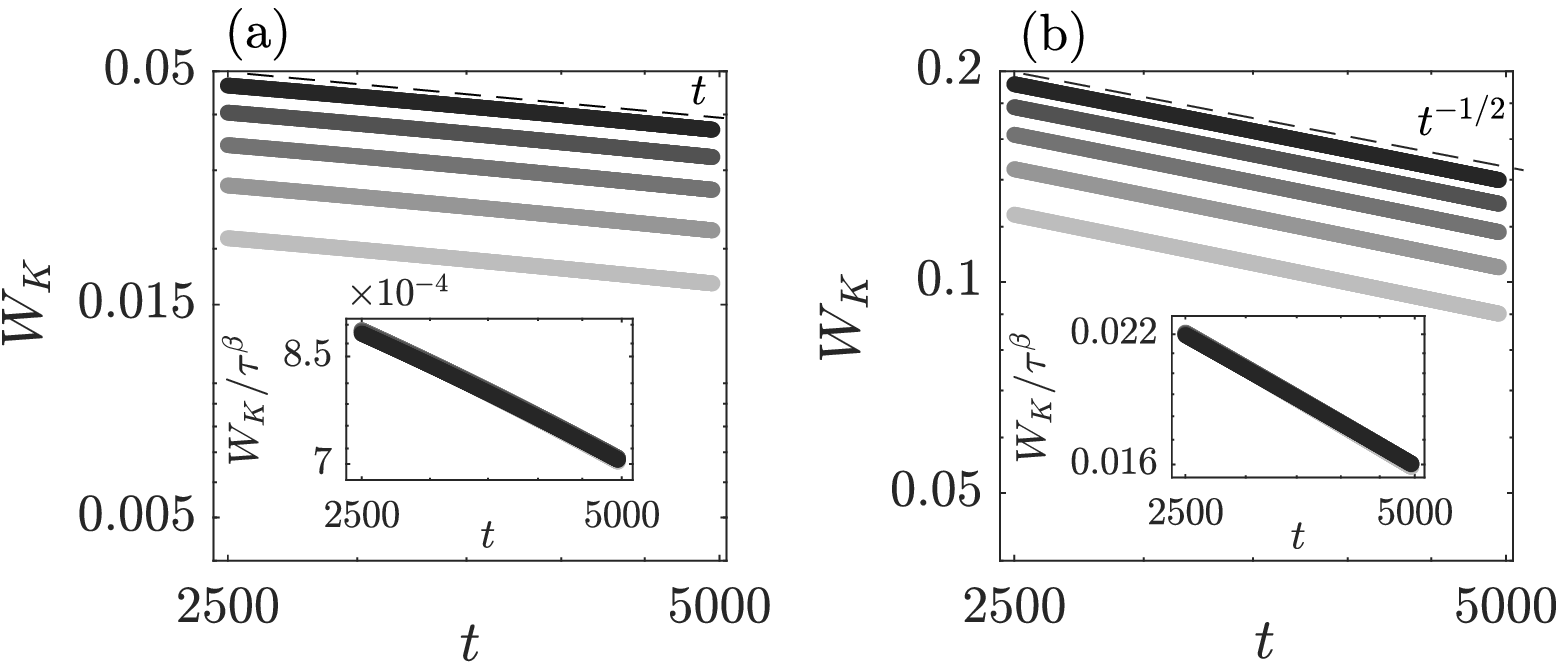}
	
	\caption{Effect of the time interval on the global scaling exponents.
		(a) Global width $W_K$ for A$\rightleftharpoons$B as a function of $t$ with $t_0=100< t < T_\tau$.
		Inset: Scaling $W_K/\tau^{\beta}$ with $\beta=1$ gives good data collapse.
		(b) Global width $W_K$ for $t_0=0.5\leq t\leq T_\tau$.
		Inset: Scaling $W_K/\tau^{\beta}$ with $\beta=1/2$ gives good data collapse.}
	
	\label{SMfig5}
\end{suppfigure}

\end{document}